\documentclass[12pt]{article}
\usepackage{amssymb,amsmath,amsfonts,amstext,graphics,graphicx}

\newtheorem{theorem}{Theorem}[section]
\newtheorem{lemma}[theorem]{Lemma}

\newtheorem{corollary}[theorem]{Corollary}
\newenvironment{proof}[1][Proof]{\begin{trivlist}
\item[\hskip \labelsep {\bfseries #1}]}{\end{trivlist}}
\newenvironment{definition}[1][Definition]{\begin{trivlist}
\item[\hskip \labelsep {\bfseries #1}]}{\end{trivlist}}

\newcommand{\qqed}{\nobreak \ifvmode \relax \else
\ifdim\lastskip<1.5em \hskip-\lastskip
\hskip1.5em plus0em minus0.5em \fi \nobreak
\vrule height0.75em width0.5em depth0.25em\fi}

\def\sgn{{\rm sgn}}

 \def\Xint#1{\mathchoice
{\XXint\displaystyle\textstyle{#1}}%
{\XXint\textstyle\scriptstyle{#1}}%
{\XXint\scriptstyle\scriptscriptstyle{#1}}%
{\XXint\scriptscriptstyle\scriptscriptstyle{#1}}%
\!\!\int}
\def\XXint#1#2#3{{\setbox0=\hbox{$#1{#2#3}{\int}$ }
\vcenter{\hbox{$#2#3$ }}\kern-.5\wd0}}

\def\dashint{\Xint-}

\def\calb{\mathcal{B}}
\def\calc{\mathcal{C}}
\def\cald{\mathcal{D}}
\def\cale{\mathcal{E}}
\def\calf{\mathcal{F}}

\def\calh{\mathcal{H}}

\def\calo{\mathcal{O}}
\def\calp{\mathcal{P}}

\def\qqed{ $\Box$}
\def\R{\mathbb{R}}
\def\C{\mathbb{C}}
\def\N{\mathbb{N}}

\def\E{{\rm I\kern-.1567em E}}
\def\A{{\rm I\kern-.1567em A}}
\def\P{{\rm I\kern-.1567em P}}
\def\V{{\rm I\kern-.1567em V}}
\def\bq{\begin{equation}}
\def\eq{\end{equation}}
\def\bqy{\begin{eqnarray}}
\def\eqy{\end{eqnarray}}

\def\del{\delta}

\def\vep{\varepsilon}

\def\ga{\gamma}

\def\la{\lambda}

\def\om{\omega}

\def\si{\sigma}

\def\varep{\varepsilon}
\def\ze{\zeta}

\def\p{\partial}

\begin{document}

\title{On Krein-like theorems for noncanonical Hamiltonian systems with continuous spectra:  application to Vlasov-Poisson}
\author{George I. Hagstrom and Philip  J. Morrison\\\\
Department of Physics and Institute for Fusion Studies\\ 
The University of Texas at Austin, Austin, TX 78712.}

\maketitle
\begin{abstract}
The notions of spectral stability and the spectrum for the Vlasov-Poisson system linearized about homogeneous equilibria,  $f_0(v)$,  are reviewed.   Structural stability is reviewed and applied to perturbations of  the linearized Vlasov operator through perturbations of  $f_0$.  We prove that for each $f_0$ there is
an arbitrarily small $\delta f_0'$ in  $W^{1,1}(\R)$ such that $f_0+\delta f_0$ is
unstable. When $f_0$ is perturbed by an area preserving rearrangement,  $f_0$ will always be
stable if the continuous spectrum is only of positive signature, where the signature of the continuous spectrum is defined as in \cite{MP92,morrison00}. If there is a signature change, 
then there is a rearrangement of $f_0$ that is unstable and arbitrarily close to $f_0$ with $f_0'$ in $W^{1,1}$.
This result is analogous to Krein's theorem for the continuous spectrum. We prove that if a discrete mode embedded 
in the continuous spectrum is surrounded by the opposite signature there is an infinitesimal perturbation
in $C^n$ norm that makes $f_0$ unstable. If $f_0$ is stable we prove that the signature of every discrete
mode is the opposite of the continuum surrounding it. 

 \end{abstract}

%\date{\today}

 %%%%%%%%%%%%%%%%%%%%%%%%%%%%%%%%%

%%%%%%%%%%%%%%%%%%%%%%%%%%%%%%%%%
 
\section{Introduction}
\label{intro}

 The perturbation of point spectra for classical vibration and quantum mechanical problems has a long history
 \cite{rayleigh,rellich}.  The more difficult problem of assessing the structural stability of the continuous spectrum
 in e.g.\ scattering problems has also been widely investigated  \cite{friedrichs,kato}.  Because general linear Hamiltonian
 systems are not governed by Hermitian or symmetric operators, the spectrum need not be stable and a transition to
 instability is possible.  For finite degree-of-freedom Hamiltonian systems, the situation is described by Krein's theorem
\cite{krein50,KJ80,moser58}, which states that a necessary condition for a bifurcation to instability
 under perturbation is to have a collision between eigenvalues of opposite signature.  The purpose of the present paper is to
 investigate Krein-like phenomena in Hamiltonian systems with a continuous spectrum. Of interest are systems that describe
 continuous media which are Hamiltonian in terms of noncanonical Poisson brackets (see e.g.\ \cite{morrison98,morrison05}). 
 
 Our study differs from that of  \cite{grillakis}, which considered canonical  Hamiltonian systems with continuous spectra in
 a Hilbert space where the time evolution operator is self-adjoint.  The effects of relatively compact
perturbations on such a system were studied and it was proved that the existence of a negative energy mode in the continuous 
spectrum caused the system to be structurally unstable.  It was also  proved that such systems are otherwise structurally 
stable.  Also, our study differs from analyses of fluid theories concerning point spectra \cite{mackay,KM95} and point and continuous spectra  \cite{hirota}, the latter using hyperfunction theory.  

 A representative example of the kind of Hamiltonian system of interest is the Vlasov-Poisson equation \cite{morrison80},
 which when linearized about stable homogeneous equilibria gives rise to a linear Hamiltonian system with pure continuous
 spectra that can be brought into action-angle form \cite{MP92,morrison94,MS94,morrison00}. A definition of signature
 was given in these works for the continuous spectrum.  In the present paper we concentrate on the Vlasov-Poisson  equation,
 but the same structure is possessed by Euler's equation for the two-dimensional fluid, where signature for shear flow
 continuous spectra was defined \cite{BM98,BM02}  and, indeed, a large class of systems \cite{morrison03}. Thus, modulo
 technicalities, the behavior treated here is expected to cover a large class of systems.

In Sec.~\ref{ncham} we review on a formal level the noncanonical Hamiltonian structure for a class of systems that includes 
the Vlasov-Poisson equation as a special case.  Linearization about  equilibria is described,  the concept of dynamical 
accessibility,   and  the linear Hamiltonian operator $T$,  the main subject of the remainder of the paper,  are defined. 
In the remainder of the paper we sketch proofs, in varying levels of detail,  pertaining to properties of this linear operator for various equilibria.  In 
Sec.~\ref{specstab} we describe spectral stability in general terms and analyze the spectrum of $T$ for the Vlasov case. 
The existence of a continuous component to the spectrum is demonstrated and  Penrose plots are used to describe the point  component.  In Sec.~\ref{structstab} we describe  structural stability and,  in particular, consider the structural stability
of $T$ under  perturbation of the equilibrium state. We show that any equilibrium is unstable under perturbation of an
arbitrarily small function in $W^{1,1}$. In Sec.~\ref{secKM} we introduce the Krein-Moser theorem and restrict to dynamically 
accessible perturbations. We prove that 
equilibria without signature changes are structurally stable and those with changes are structurally unstable. In Sec.~\ref{kreinvp}
we define critical states of the linearized Vlasov equation that are structurally unstable under perturbations that are
further restricted. We prove that a mode with the opposite signature of the continuum is structurally unstable and that 
the opposite combination cannot exist unless the system is already unstable.   Finally, in Sec.~\ref{conclu},  we conclude.

%titschmarsh, hilbert two vols by King zygmund, ... 

 %%%%%%%%%%%%%%%%%%%%%%%%%%%%%%%%%

%%%%%%%%%%%%%%%%%%%%%%%%%%%%%%%%%

\section{Noncanonical Hamiltonian form}
\label{ncham}

The class of equations of interest have a single dependent variable $\zeta(x,v,t)$, such that for each time $t$, $\zeta\colon\cald\rightarrow\R$, where the particle phase space $\cald$ is some  two-dimensional domain with coordinates $(x,v)$. The dynamics is assumed to be Hamiltonian in terms of a noncanonical Poisson bracket of the form
\bq
\{F,G\}=\int_{\cald}\!dxdv \, \zeta\,\left[\frac{\del F}{\del \zeta},\frac{\del G}{\del \zeta}\right]\,,
\label{ncze}
\eq
where  $[f,g]:=f_xg_v-f_vg_x$ is the usual Poisson bracket, where the subscripts denote partial differentiation,  and $\del F/\del \zeta$ denotes the functional derivative of a functional $F[\ze]$.  The equation of motion is generated from a Hamiltonian functional $\calh[\ze]$ as follows:
\bq
\ze_{t}=\{\ze,\calh\}=-[\ze,\cale]\,,
\eq
where $\cale:=\del \calh/\del \zeta$.  The Poisson bracket of  (\ref{ncze}) is noncanonical:  it uses only a single noncanonical variable $\ze$, instead of the usual  canonically conjugate pair;  it  possesses degeneracy reflected in the existence of Casimir invariants, $C=\int_{\cald}\!dxdv\, \calc(\ze)$, that satisfy $\{F,C\}=0$ for all functionals $F$;  but it does satisfy the Lie-algebraic properties of usual Poisson brackets.  For further details see \cite{morrison82,holm85,morrison98,morrison03}.

For the Vlasov-Poisson equation  we assume $\cald=X\times\R$,  where $X\subset\R$ or $X=S$, the circle, the distinction will not be important.  The dependent variable is the particle phase space density $f(x,v,t)$  and the Hamiltonian is given by
\bq
\calh[f]=\frac1{2}\int_{X}\!dx\!\int_{\R}\!dv\,  v^2f + \frac1{2}\int_X \!dx\,  |\phi_x|^2\,,
\eq
where $\phi$ is a shorthand for the functional dependence on $f$ obtained through solution of Poisson's equation, 
$\phi_{xx}=1-\int_{\R}\!fdv$, for  a positive charge species with a neutralizing background.  Using $\del \calh/\del f=\cale=v^2/2 +\phi$, we obtain
\bq
f_t=\{f,\calh\}=-[f,\cale]=-vf_x+\phi_x f_v \,,
\eq
where, as usual, the plasma frequency and Debye length have been used to nondimensionalize all variables.

This Hamiltonian form for the Vlasov-Poisson equation was first published in \cite{morrison80}.   
For a discussion of a general class of systems with this Hamiltonian form,  to which the ideas of the present 
analysis can be applied, see  \cite{morrison03}.   In a sequence of papers  \cite{morrison87,MP92,MS94,SM94,morrison00,MS08}
various ramifications of the Hamiltonian form have been explored -- notably,    canonization and diagonalization of the 
linear dynamics to which we now turn.

Because of the noncanonical form,   linearization requires expansion of the Poisson bracket as well as the Hamiltonian. 
Equilibria, $\ze_0$, are obtained by extremization of a free energy functional, $\calf=\calh + C$, as was first done for 
Vlasov-like equilibria in \cite{KO58}.  Writing $\ze=\ze_0 + \ze_1$ and expanding gives the Hamiltonian form for the linear
dynamics
\bq
{\ze_1}_t =\{\ze_1,H_L\}_L
\eq
where the linear Hamiltoinian, $H_L=\frac1{2}\int_{\cald}\!dxdv\,\ze_1\calo\,\ze_1$, is the second variation of $\calf$, a
quadratic from in $\ze_1$ defined by the symmetric operator $\calo$,  and $\{F,G\}_L=\int_{\cald}\!dxdv\, \ze_0[F_1,G_1]$
with $F_1:=\del F/\del \ze_1$.  Thus the linear dynamics is governed by the time evolution operator 
$T\, \cdot :=-\{\ \cdot\ ,H_L\}_L=[\ze_0,\calo\,  \cdot \, ]$.

Linearizing the Vlasov-Poisson equation   about an homogeneous equilibrium,
$f_0(v)$,  gives rise to the   system,
\bqy
{f_1}_t&=&- v{f_1}_{x} + {\phi_1}_xf_0' 
\label{f1t} \\
{\phi_1}_{xx} &=& - \int_{\R}\!dv\,  f_1\,,
\label{ph1xx}
\eqy
for the unknown $f_1(x,v,t)$.  Here $f_0':=df_0/dv$.  This is an infinite-dimensional linear Hamiltonian system
generated by the Hamiltonian functional:
\begin{align}
H_L[f_1] &= -\frac{1}{2}\int_X\!dx \int_{\mathbb{R}}\!dv \, \frac{v}{f_0'}\,  |f_1|^2+\frac{1}{2}
\int_X\!dx\, |{\phi_1}_{x}|^2\,.
\end{align}

We concentrate on systems where $x$ is an ignorable coordinate, and either Fourier expand or transform.  For Vlasov-Poisson this gives the system
\bq
{f_k}_t=- i kvf_k+\frac{if_0'}{k}  \int_{\mathbb{R}}\!d\bar{v}\, f_k(\bar{v},t)=:-T_k f_k\,,
\label{fk}
\eq
where $f_k(v,t)$  is the Fourier dual to $f_1(x,v,t)$.  Perturbation of the specturm of the  operator defined by Eq.~(\ref{fk}) is the primary subject of this paper. The operator $T_k$ is a Hamiltonian operator generated by  the  Hamiltonian functional
\begin{align}
H_L[f_k,f_{-k}] &= \frac{1}{2}\sum_k\left(-\int_{\mathbb{R}}\!dv\, \frac{v}{f_0'}\, |f_k|^2+
|\phi_k|^2\right)\,,
\end{align}
with the Poisson bracket
\begin{align}
\{F,G\}_L &= \sum_{k=1}^{\infty}ik\int_{\mathbb{R}}\!dv\, f_0'\left(\frac
{\delta F}{\delta f_k}\frac{\delta G}{\delta f_{-k}}-\frac{\delta F}{\delta f_{-k}}
\frac{\delta G}{\delta f_k}\right)\,.
\label{pbk}
\end{align}
Observe from (\ref{pbk}) that $k\in\N$ and thus  $f_k$ and $f_{-k}$ are independent variables that are almost canonically conjugate. Thus the complete system is
\bq
{f_k}_t=-T_k f_k\qquad{\rm and}\qquad {f_{-k}}_t=-T_{-k} f_{-k}\,,
\label{Tsym}
\eq
from which we conclude the spectrum is Hamiltonian.
\begin{lemma}
If $\lambda$ is an eigenvalue of the Vlasov equation linearized about the equilibrium $f_0'(v)$, then so are $-\lambda$ and $\overline{\lambda}$ (complex conjugate).  Thus if $\la=\ga +i\om$, then eigenvalues occur in the pairs, $\pm\ga$ and  $\pm i\om$, for purely  real and imaginary cases,  respectively, or quartets, $\la=\pm \ga \pm i\om$, for complex eigenvalues.
\end{lemma}
\label{specsym}
\begin{proof}
That $-\la$ is an eigenvalue follows immediately from the symmetry $T_{-k}= -T_{k}$, and that $\overline{\la}$ is an eigenvalue follows from $T_{k}f_k= -\overline{(T_{k}f_k)}$.  \qqed
\end{proof}

In \cite{MP92,MS94,morrison00,MS08}  it was shown how to scale $f_k$ and $f_{-k}$ to make them canonically conjugate variables. In order to do this requires the following definition of dynamically accessibility, a terminology introduced in \cite{MP89,MP90}.
\begin{definition}
\label{da}
A particle phase space function $k$   is {\it dynamically accessible} from a particle phase space function $h$, if $g$ is an area-preserving rearrangement of $h$; i.e.,  in coordinates $k(x,v)=h(X(x,v),V(x,v))$, where $[X,V]=1$.  A peturbation $\del h$ is {\it linearly dynamically accessible} from $h$ if $\del h=[G,h]$,  where $G$ is the infinitesimal generator of the canonical transformation $(x,v)\leftrightarrow (X,V)$.
\end{definition}

Dynamically accessible perturbations come about by perturbing the particle orbits under the action of some Hamiltonian. Since electrostatic charged particle dynamics is Hamiltonian, one can make the case that these are the only perturbations allowable within the confines of Vlasov-Poisson theory.

Given an equilibrium state $f_0$, linear dynamically accessible perturbations away from this equilibrium state satisfy  $\del f_0=[G,f_0]=G_xf_0'$.  Therefore assuming the initial condition for the linear dynamics is linearly dynamically accessible,  we can define
\bq
q_k(v,t)=f_k\qquad{\rm and}\qquad p_k(v,t)=-i{f_{-k}}/(k{f_0'})
\label{pq}
\eq
without worrying about a singularity at the zeros of $f_0'$ and $k=0$.  With the definitions of (\ref{pq}), the Poisson bracket of (\ref{pbk}) achieves canonical from
\bq
\{F,G\}_L = \sum_{k=1}^{\infty} \int_{\R}\!dv \,
\left(\frac
{\delta F}{\delta q_k}\frac{\delta G}{\delta p_k}-\frac{\delta F}{\delta p_k}
\frac{\delta G}{\delta q_k}\right)\,.
\label{cpbk}
\eq

The full system has the new Hamiltonian   $\bar \calh = \calh + U\calp$ in frame moving with speed $U$, where $\calp= \int_{\cald}\!dxdv\, vf$.  Linearizing in this frame yields the linear Hamiltonian  $\bar H_L= H_L + P_L$, from which we identify the linear momentum
\bq
P_L[f_k,f_{-k}] = \frac{1}{2}\sum^{\infty}_{k=1}\int_{\R}\!dv\, \frac{k}{f_0'}\, |f_k|^2\,,
\label{pl}
\eq
 which must be conserved by the linear dynamics. It is easy to show directly that this is the case. 
 
 \begin{lemma}
The momentum $P_L$ defined by (\ref{pl}) is a constant of motion, i.e., $\{P_L,H_L\}=0$.
\end{lemma}
\begin{proof}
This follows immediately  from (\ref{Tsym}): $\int_{\R}\!dv\,\left( f_kT_{-k} + f_{-k}T_k\right)=0$. 
 \qqed
\end{proof}

 Observe, that like the Hamiltonian, $H_L$, the momentum $P_L$ is conserved for each $k$, which in all respects appears 
 only as a parameter in our system.   Assuming the system size to be $L$ yields  $k=2\pi n/L$ with $n\in \N$, and, thus,
 this parameter can be taken to be  in $\R^+/\{0\}$. Alternatively, we could suppose $X=\R$, Fourier transform, and split the 
 Fourier integral to obtain an expression similar to (\ref{pbk}) with the sum replaced by an integral over positive values 
 of $k$.  For the present analysis we will not be concerned with issues of convergence for reconstructing the spatial 
 variation of $f_1(x,v,t)$, but only consider  $k\in \R^+/\{0\}$ to be a parameter in our operator.  
 We will see in Sec.~\ref{specstab} that the operator $T_k$ possesses a continuous component to its spectrum.  
 But, we emphasize that this continuous spectrum of interest arises form the multiplicative nature of the velocity
 operator, i.e.\ the term $vf_k$ of $T_k$, not from having an infinite  spatial  domain, as is the case for free 
 particle or scattering states in quantum mechanics.  In the remainder of the paper,  $f$ will refer to either $f_1$ 
 or $f_k$, which will be clear from context, and the dependence on $k$  will be suppressed, e.g. in $T_k$,  unless $k$
 dependence is being specifically addressed.

%%%%%%%%%%%%%%%%%%%%%%%%%%%%%%%%%

%%%%%%%%%%%%%%%%%%%%%%%%%%%%%%%%%

\section{Spectral stability}
\label{specstab}

Now we consider properties of the evolution operator $T$  defined by (\ref{fk}).  
We define spectral stability in general terms,   record some properties of $T$, and describe 
the tools necessary to characterize the spectrum of $T$.   We suppose $f_k$ varies as $\exp(-i\om t)$, 
where $\om$ is the frequency and $i\om$ is the eigenvalue.  For convenience we also use $u:=\om/k$, where 
recall $k\in \R^+$.   The system is spectrally stable if the  spectrum of $T$ is less than 
or equal to zero or the frequency is always in the closed lower half plane. Since the system  is Hamiltonian,  
the question of stability reduces to deciding if the spectrum is confined to the imaginary axis.

\begin{definition}
\label{SpectralStability}
The linearized dynamics of a Hamiltonian system around some equilibrium solution, with the 
phase space of solutions in some Banach space $\calb$,  is  {\it spectrally stable} if 
the spectrum $\sigma(T)$ of the time evolution operator $T$ is purely imaginary.
\end{definition}

Spectral stability does not guarantee that the system is stable, or that the equilibrium $f_0$ is linearly stable. (See e.g.\ \cite{morrison98} for general discussion).  The solutions of a spectrally stable system are guaranteed to grow at most sub-exponentially and  one can  construct a spectrally stable system with polynomial temporal growth for certain initial conditions. (See  e.g.\ \cite{degond86} for analysis of the Vlasov system.)

Spectral stability relies on functional analysis for its definition, since the spectrum of the operator $T$ may depend on the choice of function space $\calb$. The time evolution operators arising from the types of noncanonical Hamiltonian systems   that are of interest  here  generally contain a continuous spectrum \cite{morrison03} and the effects of perturbations that we  study can be categorized by properties of the continuous spectrum of these operators. In general for the operators of \cite{morrison03},  the operator $T$ is the sum of a multiplication  operator and an integral operator.  In the Vlasov case, the multiplicative operator is  i$v\cdot$ and the integral operator is $f_0'\int\!dv\, \cdot$.  As we will see, the multiplication 
operator causes the continuous spectrum to be composed of the entire imaginary axis except possibly for some discrete points.  

Instability comes from  the point spectrum.  In particular, the linearized Vlasov Poisson equation is not spectrally stable when   the time evolution operator has a spectrum that 
includes a point away from the imaginary axis, with the necessary counterparts  implied by  Lemma \ref{ncham}.\ref{specsym}. For the operator $T$ this will always be a discrete mode; i.e.\ an eigenmode associated with an eigenvalue in the point spectrum.

\begin{theorem}
\label{SpecTheorem}
The one-dimensional linearized Vlasov-Poisson system with homogeneous equilibrium $f_0$ is spectrally unstable 
if for some $k\in\R^+$ and $u$ in the upper half plane,  the plasma dispersion relation 
\[
\vep(k,u):=1-k^{-2}\int_{\R}\!dv 
\frac{f_0'}{v-u}=0\,.
\]
 Otherwise it is spectrally stable.
\end{theorem}
\begin{proof}
The details of this proof are given in plasma textbooks.  It follows directly from (\ref{f1t}) and (\ref{ph1xx}), and the assumption  $f_1\sim\exp(ikx-i\om t)$. 
 \qqed
\end{proof}

Using the Nyquist method  that relies on the argument principle of complex analysis,  Penrose \cite{penrose} was able to relate the vanishing of $\vep(k,u)$ to the winding number  of the closed curve determined by the real and imaginary parts of $\vep$ as $u$ runs along the real axis.  Such closed curves are called Penrose plots.  The crucial quantity is the  integral part of $\vep$ as $u$ approaches the real axis from above:  
\[
\lim_{u\rightarrow 0^+}\frac{1}{\pi}\int_{\R}\!dv\, \frac{f_0'}{v-u}=H[f_0'](u)-if_0'(u)\,,
\]
 where $H[f_0']$ denotes the Hilbert transform, $H[f_0']=\frac{1}{\pi}\dashint\!dv\,{f_0'}/({v-u})$, where $\dashint:=PV\int_{\R}$   indicates the Cauchy principle value.  (See \cite{king} for an in depth treatment of Hilbert transforms.)   The graph of the real line under this mapping is the essence of  the Penrose plot, and so we will refer to these closed curves as Penrose plots as well.  When necessary to avoid ambiguity we will refer to the former as $\vep$-plots. 

\begin{figure}[htbp]
%\begin{center}
\includegraphics[scale=.66]{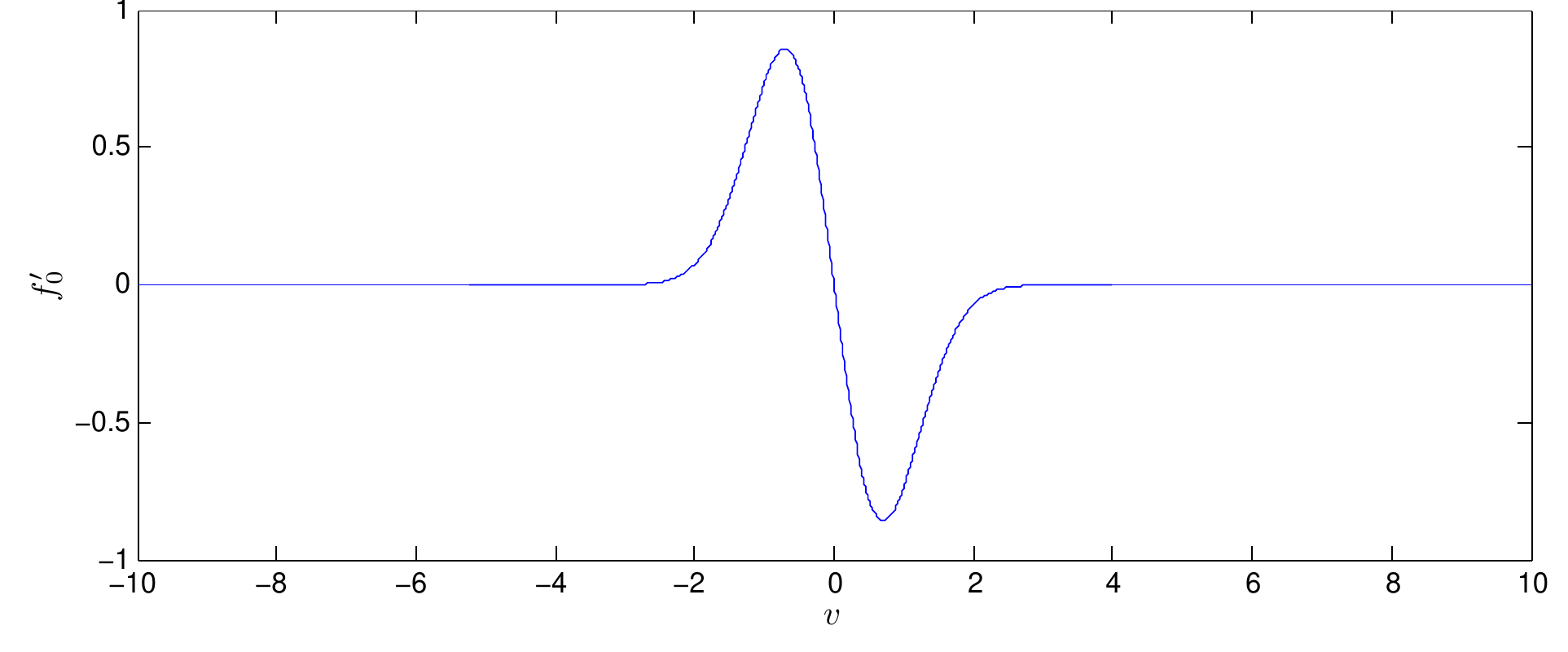}
%\includegraphics[height=4 in,width= 6in]{s1gaussian.pdf}
%\end{center}
\caption{$f_0'$ for a Maxwellian  distribution.}
\label{gaus1}
\end{figure}

\begin{figure}[htbp]
%\begin{center}
\includegraphics[scale=.6]{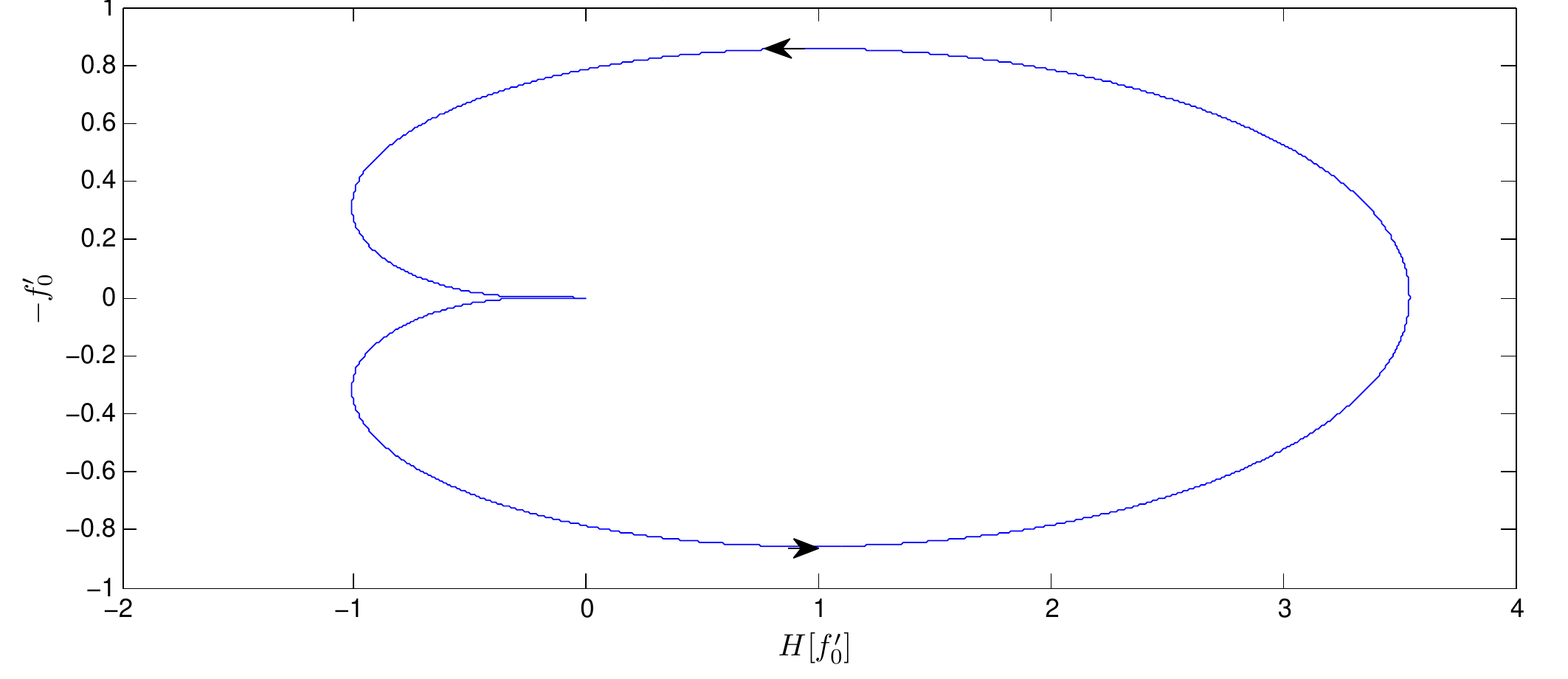}
%\end{center}
\caption{Stable Penrose plot for a Maxwellian distribution.}
\label{penroseGaus}
\end{figure}

For example, Fig.~\ref{gaus1}  shows the derivative of the distribution function, $f_0'$,   for the case of a Maxwellian distribution 
and Fig.~\ref{penroseGaus}  shows  the contour $H[f_0']-if_0'(u)$ that emerges from the origin in the complex plane at $u=-\infty$,  descends, and then wraps  around to return to the origin at $u=\infty$.  From this figure it is evident that the winding number of the $\vep(k,u)$-plot  is zero for any fixed $k\in\R$, and as a result there are no unstable modes.

Making use of the argument principle as described above,   Penrose obtained the following criterion: 
\begin{theorem}
\label{PenroseCriterion}
The linearized Vlasov-Poisson system with homogeneous equilibrium  $f_0$ is spectrally unstable if there exists a point $u$ such that 
\[
f_0'(u)=0 \qquad {\rm and}\qquad  \dashint\!dv\,  \frac{f_0'(v)}{v-u}>0\,,
\]
with $f_0'$ traversing zero at $u$.  Otherwise it is spectrally stable.
\end{theorem}

\begin{figure}[htbp]
\begin{center}
\includegraphics[scale=.6]{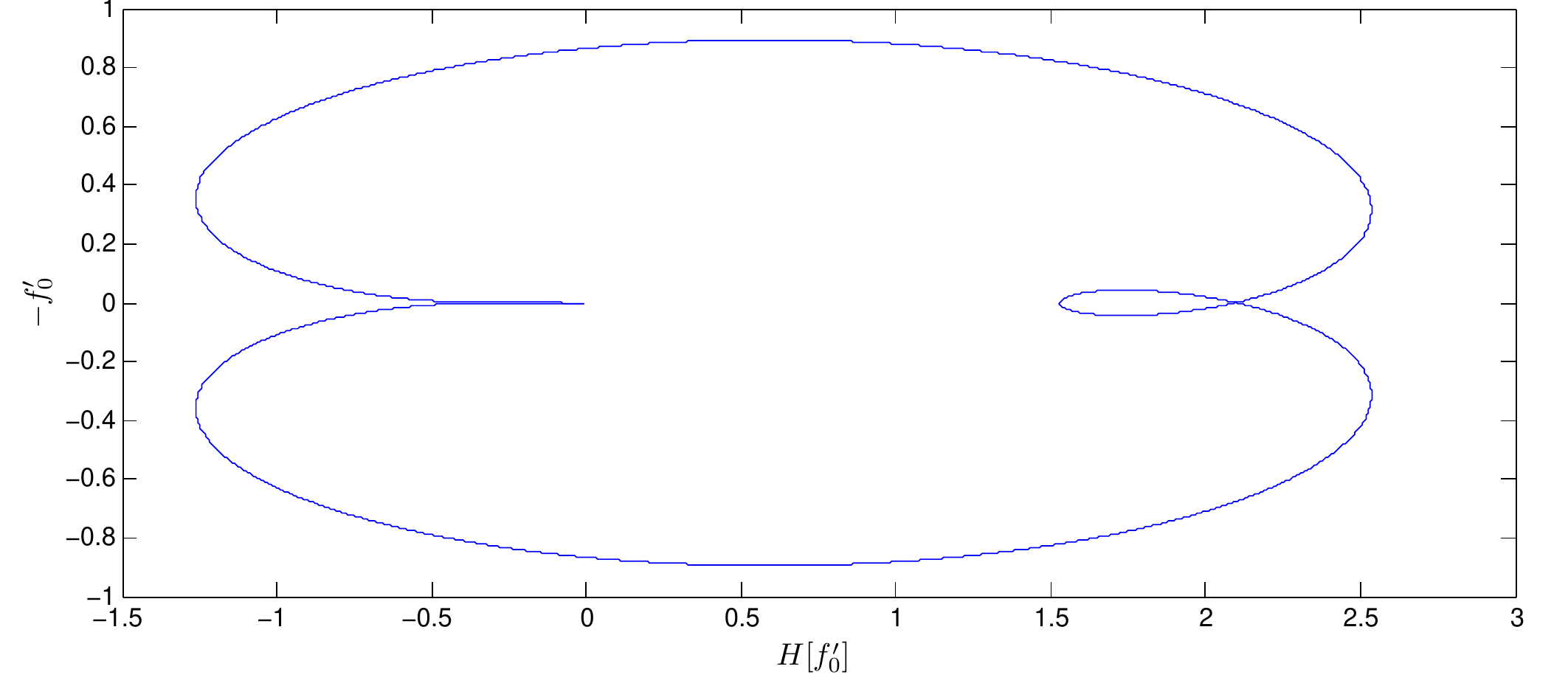}
\end{center}
\caption{Penrose plot for a stable superposition of Maxwellian distributions}
\label{2max}
\end{figure}

\begin{figure}[htbp]
\begin{center}
\includegraphics[scale=.6]{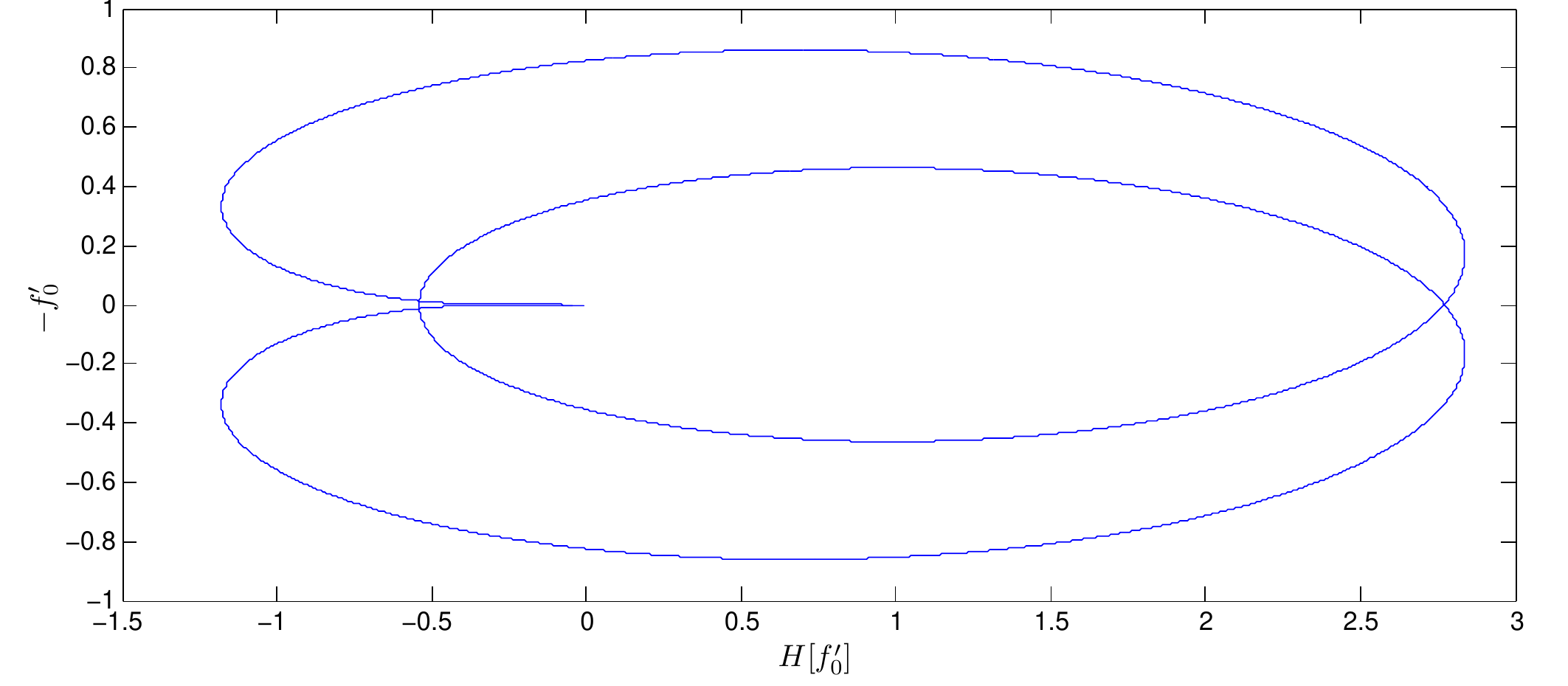}
\end{center}
\caption{The unstable Penrose plot corresponding to two separated Maxwell distributions.}
\label{uspen}
\end{figure}

Penrose plots can be used to visually determine   spectral stability.  As described above, the  Maxwellian distribution
$f_0=e^{-v^2}$ is stable, as the resulting $\varep$-plot does not encircle the origin. However, it is not difficult to 
construct unstable distribution functions. The superposition of two displaced Maxwellian distributions, 
$f_0=e^{-(v+c)^2}+e^{-(v-c)^2}$, is such a case. 
As $c$ increases the distribution goes from stable to unstable.  Figures \ref{2max} and \ref{uspen} demonstrate how the transition from stability to  instability is manifested in a Penrose plot. The two examples are $c=3/4$ and $c=1$.  
(Note, the normalization of $f_0$ only affects the overall scale of the Penrose plots and so is ignored for convenience.)  It is evident from Fig.~\ref{uspen} that for some $k\in\R$ the $\vep$-plot (which is a displacement of  the curve shown by multiplying by $-k^{-2}$ and adding unity) will encircle the origin, and thus will be unstable for such $k$-values.

%\subsection{Spectral Theory of the operator $T$}

We now are positioned to completely determine the spectrum.   For convenience we set $k=1$ when it does not affect the essence of our arguments, and consider  the operator $T\colon f\mapsto  ivf- if_0'\int f$ in the space
$W^{1,1}(\mathbb{R})$, but we also discuss the space $L^1(\mathbb{R})$.  The space $W^{1,1}(\mathbb{R})$ is the Sobolev space containing 
the closure of functions under the
norm $\|f\|_{1,1}=\|f\|_1+\|f'\|_1$. Thus it contains all functions that are in $L^1(\mathbb{R})$ whose weak 
derivatives are also in $L^1(\mathbb{R})$. First we establish the expected facts that $T$ is densely defined and closed.

In $W^{1,1}$ the operator $T$ is the sum of the multiplication operator and a bounded operator -- that it is densely defined and
closed follows from the fact that the multiplication operator is densely defined and closed in these spaces, where
\[
D_1(T):= \{f|vf\in W^{1,1}(\mathbb{R})\}.
\]

\begin{theorem}
The operator $T\colon W^{1,1}(\mathbb{R})\rightarrow W^{1,1}(\mathbb{R})$ with domain $D_1(T)$ is both 
(i) densely defined   and  (ii)   closable.
\end{theorem}
\begin{proof}
(i) The set of all smooth functions with compact support, $C_c^{\infty}(\mathbb{R})$ is a subset of $D_1$.
This set is dense in $W^{1,1}(\mathbb{R})$ so $D_1$ is dense and $T$ is densely defined.
(ii) The operator $T$ is closable if the operator $v$ is closable because $T$ and $v$ differ by a bounded operator.
The multiplication operator $v$ is closed if for each sequence $f_n\subset W^{1,1}(\mathbb{R})$ that converges to
$0$ either $vf_n$ converges to $0$ or $vf_n$ does not converge. Suppose $vf_n$ converges. At each point 
$f_n$ converges to $0$. Therefore $vf_n$ converges to $0$ at each point, so $vf_n$ converges to $0$ if it converges.
\qqed
\end{proof}

Therefore there exists some domain $D$ such that the graph $(D,TD)$ is closed.

\medskip

In determining  the spectrum of the operator $T$, denoted $\si(T)$,  we split 
the spectrum into   point, residual,  and continuous components as follows:

\begin{definition}
\label{spectra}
For  $\lambda\in\sigma(T)$  the resolvent of $T$ is  $R(T,\lambda)=(T-\lambda I)^{-1}$, where $I$ is the identity operator. We say $\lambda$
is  (i) in the point spectrum,  $\sigma_p(T)$,  if $T-\lambda I$ fails to be injective, (ii) in the
residual spectrum,   $\sigma_r(T)$, if  $R(T,\lambda)$ is not densely defined, and (iii) in the continuous spectrum,   
$\sigma_c(T)$, if $R(T,\lambda)$ is densely defined but unbounded.
\end{definition}

Using this definition we characterize the spectrum of the operator $T$.

\begin{theorem}
\label{spectrumT}
The component $\sigma_p(T)$ consists of all points $\la=iu\in \C$ where $1-k^{-2}\int_{\R}\!dv\,{f_0'}/({v-u})=0$, $\sigma_c(T)$
 consists of all $\lambda=i u$ with $u\in\mathbb{R}/(-i
\sigma_p(T)\cap\mathbb{R})$, and $\sigma_r(T)$ contains all  points $\lambda=iu$ in the complement of $\sigma_p(T)\cup \sigma_c(T)$ that satisfy $f_0'(u)=0$.
\end{theorem}
\begin{proof}
By the Penrose criterion we can identify all the points in the
point spectrum. If
$1-k^{-2}\int_{\R} \,dv{f_0'}/({v-u})=0$ then $iu=\lambda\in\sigma_p(T)$.
Because the system is Hamiltonian these modes will occur for the linearized Vlasov-Poisson system in quartets (two for $T_k$ and two for $T_{-k}$), as follows from Lemma \ref{ncham}.\ref{specsym}.
It is possible for there to be discrete modes with real frequencies and these will occur in pairs.  If
for real $u$ the  map $u\mapsto \varep$ passes through the origin then there will be such an embedded  mode.

For convenience we drop the wavenumber subscript  $k$ on $f_k$ and add the subscript $n$ to identify $f_n$ as an element of a  sequence of functions that converges to zero with,   for each $n$,  support contained
in an interval of length $2\epsilon(n)$ surrounding the point $u$  and zero average value. Let $u\in\mathbb{R}$ and  choose  the sequence $\{f_n\}$  so that $\epsilon(n)\rightarrow 0$. Then for each $n$
\begin{align}
\|R(T, iu)\| &\geq \frac{\|f_n\|_{1,1}}{\|(v-u)f_n\|_{1,1}}\nonumber \\
&\geq \frac{\|f_n\|_{1,1}}{\|v-u\|_{W^{1,1}(u-\epsilon,u+\epsilon)}\|f_n\|_{1,1}} \nonumber\\
&=\frac{1}{\|v-u\|_{W^{1,1}(u-\epsilon,u+\epsilon)}}
\nonumber
\end{align}
In the above expression,  $W^{1,1}(u-\epsilon,u+\epsilon)$ refers to the integral of $|f|+|f'|$ over the interval $(u-\epsilon,u+\epsilon)$.
Therefore the resolvent is an unbounded operator and $iu=\lambda$ is in the spectrum. If the frequency $u$ has an imaginary component $i\gamma$ then
$\|R(T,iu)\|<1/\gamma$ so unless $iu=\lambda$ is part of the point spectrum it is part of the resolvent set.

The residual spectrum of $T$ is contained in the point spectrum of $T^*$. The dual of $W^{1,1}$ is the space $W^{-1,1}$ defined by pairs $(g,h)\in W^{-1,1}$ with
$\|(g,h)\|_{{-1,1}}<\infty$ (cf.\  \cite{adams}). The operator $T^*(g,h)= i(vg-h+\int (gf_0'-hf_0'')dv,-vh)$ is the adjoint of $T$. If we search for a
member $iu=\lambda$ of the point spectrum we get two equations, one of which is $(v-u)h=0$. This forces $h=0$ because $h$ cannot be a $\delta$-function in $W^{-1,1}$.
The other equation is then $(v- u)g+\int gf_0'dv=0$ which
can only be true if the integral is zero or if $(v- u)g$ is a constant. For this $g=\frac{1}{v- u}$ and the resulting equation for $u$ is the
same equation as that for the frequency of the point modes of $T$. If the integral is zero then $g=\delta(v-u)$ is a solution when $f_0'(u)=0$. Therefore the residual
spectrum contains the points $\lambda=iu$ satisfying $f_0'(u)=0$.
\qqed
\end{proof}

This characterization of the spectrum fails in Banach spaces with less regularity than $W^{-1,1}$, such as $L^p$ spaces, because the Dirac $\delta$ is not contained in the dual space. In this case  the
residual spectrum vanishes because $\sigma_p(T^*)=\sigma_p(T)$. This calculation is nearly identical to that of Degond \cite{degond86}, who characterizes the residual spectrum slightly differently than
we do.  In any event, the result is that the Penrose criterion determines whether $T$ is spectrally stable or not. If the winding number of the $\varep$-plot is positive,  then there is spectral instability and if it
is zero there is   spectral stability.

%%%%%%%%%%%%%%%%%%%%%%%%%%%%%%%%%

%%%%%%%%%%%%%%%%%%%%%%%%%%%%%%%%%

\section{Structural stability}
\label{structstab}

Spectral stability characterizes the linear dynamics of a nonlinear Hamiltonian system
in a neighborhood of an equilibrium.  The main  question now is to determine when a spectrally
stable system can be made spectrally unstable with a small perturbation. When this is impossible for our choice of allowed perturbations, 
we say the equilibrium is structurally stable, and when there is an infinitesimal perturbation
that makes the system spectrally unstable we say that the equilibrium is spectrally unstable.
We can make this more precise by stating it in terms of operators on a Banach space.

\begin{definition}
\label{genstructuralstability}
Consider an equilibrium solution of a Hamiltonian system and the corresponding 
time evolution operator $T$ for the linearized dynamics, with a  phase space  some
Banach space $\calb$. Suppose that $T$ is spectrally stable. Consider perturbations 
 $\delta T$ of $T$ and define a norm on the space of such perturbations. Then we say that the 
equilibrium is {\it structurally stable} under this norm if there
is some $\del>0$ such that for every $\|\delta T\|<\del$ the operator $T+\delta T$ is spectrally stable. 
Otherwise the system is  {\it  structurally unstable}.
\end{definition}

Because we are dealing with physical systems
it makes sense to have physical motivation for the choice of norm on the space of 
perturbations. In this paper we
are interested in perturbations of the Vlasov equation through changes in the equilibrium. This
choice is motivated by the Hamiltonian structure of the equations and  Krein's theorem for finite-dimensional systems. In general the space
of possible perturbations is quite large, but perturbations of equilibria give rise to operators in certain Banach spaces and motivate the definition of norm.    Even in the case of unbounded perturbations there may exist such a  norm 
(see Kato \cite{kato}, for instance).

Consider a stable equilibrium function $f_0$. We will consider perturbations of the 
equilibrium function and the resulting perturbation of the time evolution operator. Suppose that the time
evolution operator of the perturbed system is $T+\delta T$.
In the function space that we will consider these perturbations are bounded operators 
and their size can be measured by the norm $\|\delta T\|$. This norm will be proportional
to the norm of $\|\delta f_0'\|$,  where $\delta f_0$ is the
perturbation of the equilibrium.

\begin{definition}
\label{StructuralStability}
Consider the formulation of the linearized Vlasov-Poisson equation in the Banach
space $W^{1,1}(\mathbb{R})$ with a spectrally stable homogeneous equilibrium function $f_0$. Let $T_{f_0+\delta f_0}$ 
be the time evolution operator corresponding to the linearized dynamics around the distribution 
function $f_0+\delta f_0$. If there exists some $\del$ depending only on $f_0$ such
that $T_{f_0+\delta f_0}$ is spectrally stable whenever $\|\delta T_{\delta f_0}\|=\|T_{f_0}-T_{f_0+\delta f_0}\|<\del$,  then the equilibrium $f_0$
is structurally stable under perturbations of $f_0$. 
\end{definition}

The aim of this work is  to characterize the structural stability of the linearized
Vlasov-Poisson equation. 
We will prove that if the perturbation function is some homogeneous $\delta f_0$ and the norm
is $W^{1,1}$ (and $L^1$ as a consequence) every equilibrium distribution function is structurally unstable
to an infinitesimal perturbation in this space. This fact will force us to consider more restricted sets of 
perturbations.

\subsection{Winding number}

We need to compute the winding number of Penrose plots and the change in winding number 
under a perturbation,  both in this section and the rest of the paper. 
%The method
%we find most effective for understanding the winding number comes from differential topology.
We use the fact that one way to compute the winding number is to draw a ray from the origin to infinity and
to count the number of intersections with the contour accounting for  orientation. 
 \begin{lemma}
\label{winding}
Consider an equilibrium distribution function $f_0'$. The winding number of the
Penrose  $\varep$-plot around the origin is equal to $\sum_u \sgn(f_0''(u))$ for all 
$u\in\mathbb{R}^-$,  satisfying $f_0'(u)=0$.
\end{lemma}

To calculate the winding number of the Penrose $\vep$-plot using this lemma one counts the number of zeros of
$f_0'$ on the negative real line and adds them with a positive sign if $f_0''$ is positive,   
a Penrose  crossing from the upper  half plane to the lower half plane, a negative sign if
$f_0''$ is negative,    a crossing from the lower half plane to the upper
half plane, and zero if $u$ is not a crossing of the x-axis, a tangency. 
This lemma comes from the following equivalent characterization of the
winding number from differential topology \cite{guillemin}.
\begin{definition}
\label{diffwinding}
If $X$ is a compact, oriented, $l$-dimensional manifold and $f\colon X\rightarrow\mathbb{R}^{l+1}$
is a smooth map, the winding number of $f$ around any point $z\in\mathbb{R}^{l+1}-f(X)$
is the degree of the direction map $u\colon X\rightarrow S^l$ given by $u(x)=\frac{f(x)-z}{|f(x)-z|}$.
\end{definition}

In our case the compact manifold is the real line plus the point at $\infty$ and $l=1$. 
The degree of $u$ is the intersection number of $u$ with any point on the circle taken
with a plus sign if the differential preserves orientation and a minus sign if it reverses
it. The lemma is just a specialization of this definition to the negative $x$-direction on the circle. If more than one derivative of $f_0$ vanishes at a zero of $f_0'$ there is a standard procedure for calculating the winding number by determining if there is a sign change in
$f_0'$ at the zero.

\subsection{Structural instability of general $f_0$}

In a large class of function spaces it is possible to
create infinitesimal perturbations that make any equilibrium distribution function unstable. This can happen
in any space where the Hilbert transform is an unbounded operator. In these spaces
there will be an infinitesimal $\delta f_0$ such that $H[\delta f_0']$ is order one at a zero of $f_0'$. Such a 
perturbation can turn any point where $f_0'=0$ into a point where $H[f_0'+\delta f_0']>0$
as well. Because $\delta f_0'$ is small and the region where $H[\delta f_0']$ is not small is also small,  the only
effect on the Penrose plot will be to move the location of the zero. Thus, such a perturbation will increase the winding number and cause instability.

We will explicitly demonstrate this for the Banach space $W^{1,1}(\R)$ and, 
by extension, the Banach space $L^1\cap C_0$. This will imply that any distribution 
function is infinitesimally close to instability when the problem is set in one 
of these spaces, implying the structural instability of every distribution function.

Suppose we perturb $f_0$ by a function $\delta f_0$. The resulting perturbation to
the operator $T$ is the operator mapping $f$ to $\delta f_0'\int dv f$. In the space
$W^{1,1}$ this is a bounded operator and thus we take the norm of the perturbing operator
to be $\|\delta f_0'\|_{1,1}$. Now we introduce a class of perturbations that can be
made infinitesimal,  but have Hilbert transform of order unity.

\begin{figure}[htbp]
\begin{center}
\includegraphics[scale=.65]{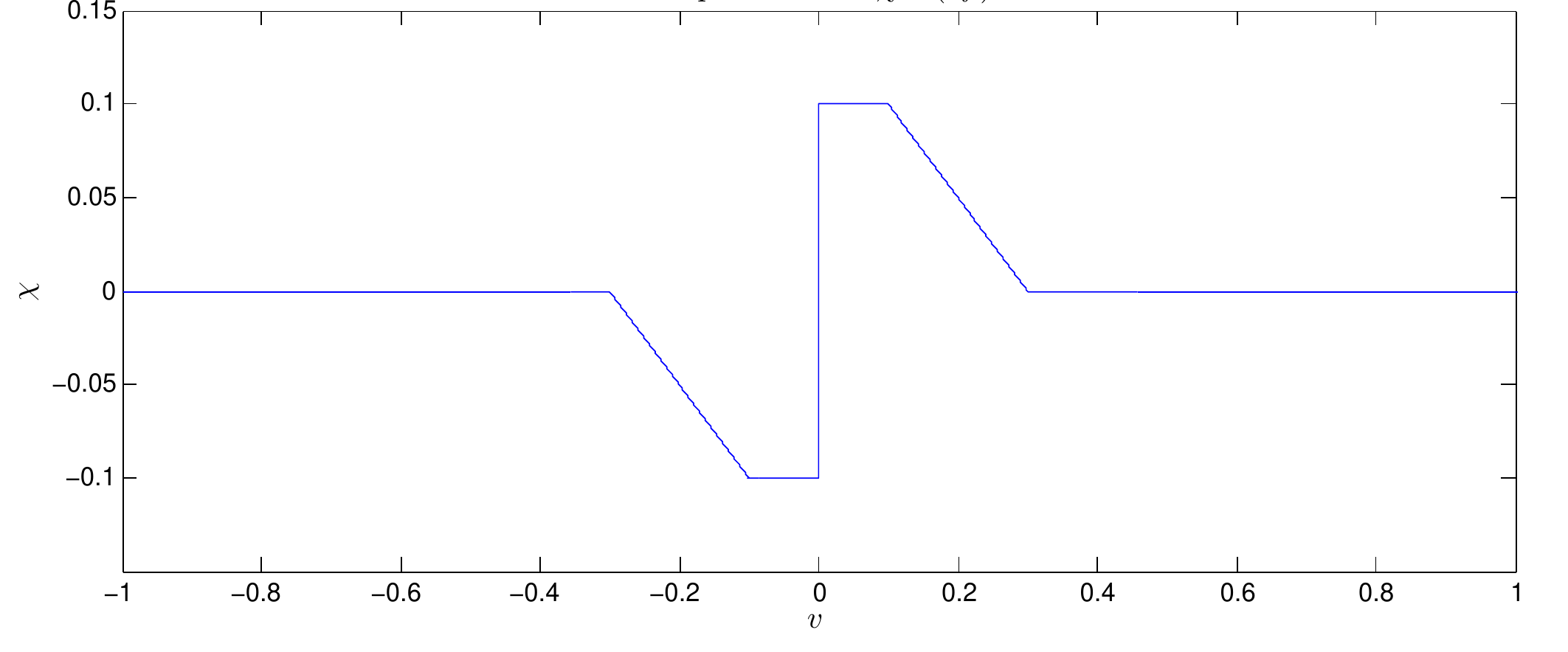}
\end{center}
\caption{The perturbation $\chi$ for $\epsilon=e^{-10}$, $h=d=.1$.}
\label{chi}
\end{figure}

\begin{figure}[htbp]
\begin{center}
\includegraphics[scale=.85]{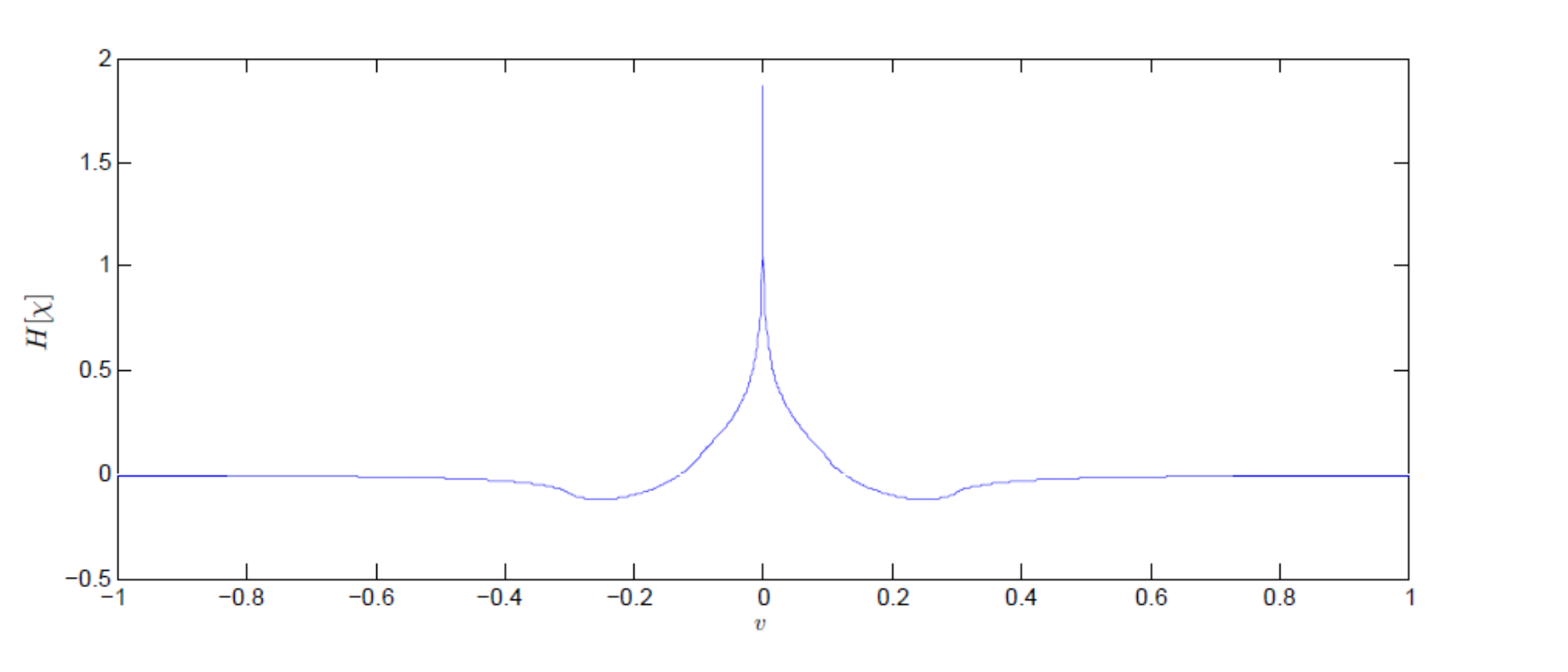}
\end{center}
\caption{The Hilbert transform of $\chi$.}
\label{Hchi}
\end{figure}

Consider the function 
$\chi(v,h,d,\epsilon)$ defined by 
\[
\chi = 
\left\{
\begin{array}{ll}
{hv}/{\epsilon} &\quad  |v|<\epsilon \\
  h\, \sgn(v) &\quad \epsilon<|v|<d+\epsilon \\
  h+d/2+\epsilon/2-v/2 &\quad 2h+d+\epsilon>v>d+\epsilon \\
  -h-d/2-\epsilon/2-v/2 &\quad 2h+d+\epsilon>-v>d+\epsilon \\
  0 &\quad |v|>2h+d+\epsilon
 \end{array}\,.
 \right.
 \]
 Figures \ref{chi} and \ref{Hchi}  show the graph of $\chi$ and its Hilbert transform, $H[\chi]$, respectively.

\bigskip

\begin{lemma}
 \label{chi_lemma}
 If we choose $d=h$ and $\epsilon=e^{-(1/h)}$, then  for any $\delta,\gamma>0$ we can choose
 an $h$ such that $\|\chi\|_{1,1}<\delta$ and $\dashint\!dv\, {\chi}/{v}>1-O(h)$, 
 and $|\dashint\!dv\, {\chi}/({u-v})|<|\gamma/u|$ for $|u|>|2h+d+\epsilon|$.
 \end{lemma}
\begin{proof}
In the space $W^{1,1}$ the function $\chi$ has norm $2h^2+2hd+h\epsilon + 4h$, which is less than any  $\delta$ for small enough $h$. 
  We can compute the value of the Hilbert transform of this 
 function at a given point $u$ by calculating the principal values:
 \bqy
  \dashint\!dv\, \frac{\chi}{v-u}&=& \frac{hu}{\epsilon}\log\left(\frac{|u-\epsilon|}{|u+\epsilon|}\right)
+h\log\left(\frac{|d+\epsilon-u||u+d+\epsilon|}{|\epsilon+u||\epsilon-u|}\right)\nonumber\\
&+&   \frac{1}{2}(d+\epsilon+2h-u)\log\left(\frac{|d+\epsilon+2h-u|}{|d+\epsilon-u|}\right)
\nonumber\\
&+&
 \frac{1}{2}(d+\epsilon+2h+u)\log\left(\frac{|d+\epsilon+2h+u|}{|d+\epsilon+u|}\right)\,.
 \label{chiint}
\eqy
 We analyze the asymptotics of this function as $h$, $d$, and
$\epsilon$ go to zero, with the desiderata that i) the norm of $\chi$
goes to zero,  ii) the maximum of the Hilbert transform of $\chi$ is $O(1)$,  and iii) there
is a band of vanishing width around the origin outside of which the Hilbert 
transform can be made arbitrarily close to zero. 

Note that (\ref{chiint})  can be written as a linear combination of translates of
the function $x\log x$:
\bqy
\dashint\!dv\, \frac{\chi}{v-u} &=&
 \frac{h}{\epsilon}((u-\epsilon)\log(|u-\epsilon|)
-  (u+\epsilon)\log(|u+\epsilon|))
\nonumber\\
&-& \frac{1}{2}(d+u+\epsilon)\log(|d+u+\epsilon|) -\frac{1}{2}(d-u+\epsilon) \log(|d-u+\epsilon|)
\nonumber\\
&+&\frac{1}{2}(d+u+\epsilon+ 2h)\log(|d+u+\epsilon+2h|)
\nonumber\\
&+&\frac{1}{2}(d-u+\epsilon+2h) \log(|d-u+\epsilon+2h|)\,.
\label{chiagain}
\eqy
The function $x\log x$ has a local minimum for positive $x$ at $x=1/e$. 
This is the point at which the function is most negative. It has zeros at $x=0$
and $x=1$. For values of $u,d,\epsilon,h$ close to zero all of the arguments 
of the $\log$ functions are less than $1/e$. Therefore,   for $|u|<d+\epsilon+2h$ 
the $x\log x$ terms are all monotonically decreasing functions of the 
argument $x$. Of the terms of (\ref{chiagain}),    $\frac{h}{\epsilon}((u-\epsilon)\log(|u-\epsilon|)
-(u+\epsilon)\log(|u+\epsilon|))$ has by far the largest coefficient as long as
$\epsilon$ is much smaller than $h$. We choose $h=d$ and $\epsilon=0(e^{-1/h})$. Then the terms that
do not involve $\epsilon$ are all smaller than $(6h+\epsilon)\log (6h+\epsilon)$. With these choices $\chi$ satisfies
\bqy
\chi(0)&=&2-(h+e^{-1/h})\log(|h+e^{-1/h}|)+(3h+e^{-1/h})\log(|3h+e^{-1/h}|)\nonumber\\
&=&2+O(h\log h)\,.
\nonumber
\eqy
Consider the pair of functions $-(u+c)\log(|u+c|)+(u-c)\log(|u-c|)$. The derivative
with respect to $u$ is $-\log(|u+c|)+\log(|u-c|)$. This is zero for $u=0$ and for $u>0$ it is always negative and the pair is
always decreasing, and for small values of $h$ the pair is guaranteed to be positive. Suppose that $u>\epsilon$.
Then we can bound the term with the $h/\epsilon$ coefficient:
\bqy
&{\ }&\frac{h}{\epsilon}\left|(u-\epsilon)\log(|u-\epsilon|)-(u+\epsilon)\log(|u+\epsilon|)\right|  \nonumber\\
&{\ }& \hspace{1.0 in} =\left|\frac{h}{\epsilon}(u-\epsilon)\log\frac{|u-\epsilon|}{|u+\epsilon|}-2\epsilon\log(|u+\epsilon|)\right| \nonumber\\
&{\ }& \hspace{1.0 in} = \frac{h}{\epsilon}\left|(u-\epsilon)\log\frac{1-\frac{\epsilon}{u}}{1+\frac{\epsilon}{u}}-2\epsilon\log(|u+\epsilon|)\right|\nonumber \\
&{\ }& \hspace{1.0 in}  < \frac{h}{\epsilon}\left|(u-\epsilon)\log(e^{-\epsilon/u})\right|+2\left|h\log(|u+\epsilon|)\right| \nonumber\\
&{\ }&\hspace{1.0 in}  = \frac{h(u-\epsilon)}{u}+2\left|h\log(|u+\epsilon|)\right|\,.
\nonumber
\eqy
For $u>>\epsilon$, for example if $u=O(h^2)$,   this term is  $O(h\log h)$. Therefore,  for $|u|>h^2$ we have $\chi=O(h\log h)$ which
can be made arbitrarily small. When $|u|>3h+\epsilon$ the function $\chi$ decreases at least as fast as $O(1/u)$. With these choices of
$h$, $d$, and $\epsilon$,  the norm of $\chi$ is $O(h)$, which proves the Lemma.
 \qqed
\end{proof}

Now we state the theorem that any equilibrium is strucutrally unstable in both 
the spaces $W^{1,1}$ and $L^1\cap C_0$.

\begin{theorem}
\label{structuralinstability}
A stable equilibrium distribution $f_0\in C^2$ is structurally unstable under perturbations of the equilibrium in the Banach spaces $W^{1,1}$ and    $L^1\cap C_0$. 
\end{theorem}
\begin{proof}

If $f_0$ is stable then the Penrose $\epsilon$-plot of $f_0'$ has a winding number of zero. Because the point at $\infty$ corresponds
to a crossing where $f_0'$ goes from negative to positive there exists a point $u_0$ with $f_0'(u_0)=0$  that is an isolated zero, 
$H[f_0'](u_0)<0$, and $f_0''(u_0)<0$. Let $F=\sup |f_0''|$. Choose $h$ to always be smaller than the distance from $u_0$ to
the nearest $0$ of $f_0'$. Then if $\epsilon=O(e^{-1/h})$ and $d=h$ the support of $\chi(u-u_0)$ will contain only one zero of $f_0'$.
For $h$ small enough the slope of $\chi$ at $u_0$ will be greater than $F$ so that the function $f_0'+\chi$ will be positive for $u$ 
in the set $(u_0,u^+)$ for some $u^+$ in the support of $\chi$. Similarly $f_0'+\chi$ will be negative for $u$ in the set $(u^-,u_0)$
for some $u^-$ in the support of $\chi$. Because $\chi$ has compact support the function $f_0'+\chi$ is positive in a neighborhood
outside of the support of $\chi$ so that the intermediate value theorem guarantees one additional zero of the function $f_0'+\chi$ for
$u>u_0$ and also for $u<u_0$. Choose $\chi$ so that this Hilbert transform of $f_0'+\chi$ is positive at the point $u_0$
and $h$ small enough that it is negative before the next zero of $f_0'+\chi$ on either side of $u_0$. Then the winding number of
$f_0'+\chi$ is positive because an additional positive crossing has been added on the negative real line. 

Because the norm of $\chi$ is $O(h)$ in both $W^{1,1}$ and $L^1$ the distribution $f_0$ is unstable to an arbitrarily small 
perturbation and is therefore structurally unstable.
\qqed
\end{proof}

\begin{figure}[htbp]
\begin{center}
\includegraphics[scale=.6]{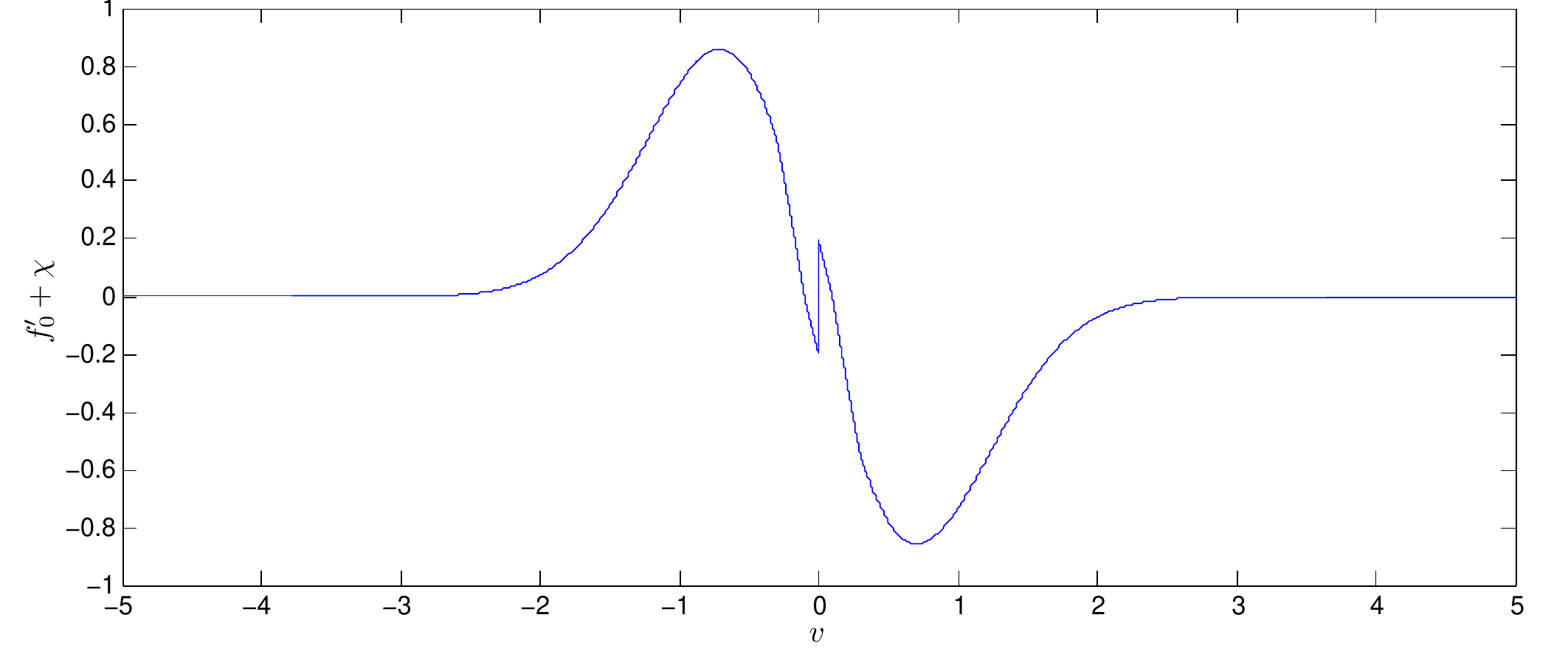}
\end{center}
\caption{$f_0'+\chi$ for a Maxwellian distribution.}
\label{pert}
\end{figure}

\begin{figure}[htbp]
\begin{center}
\includegraphics[scale=.6]{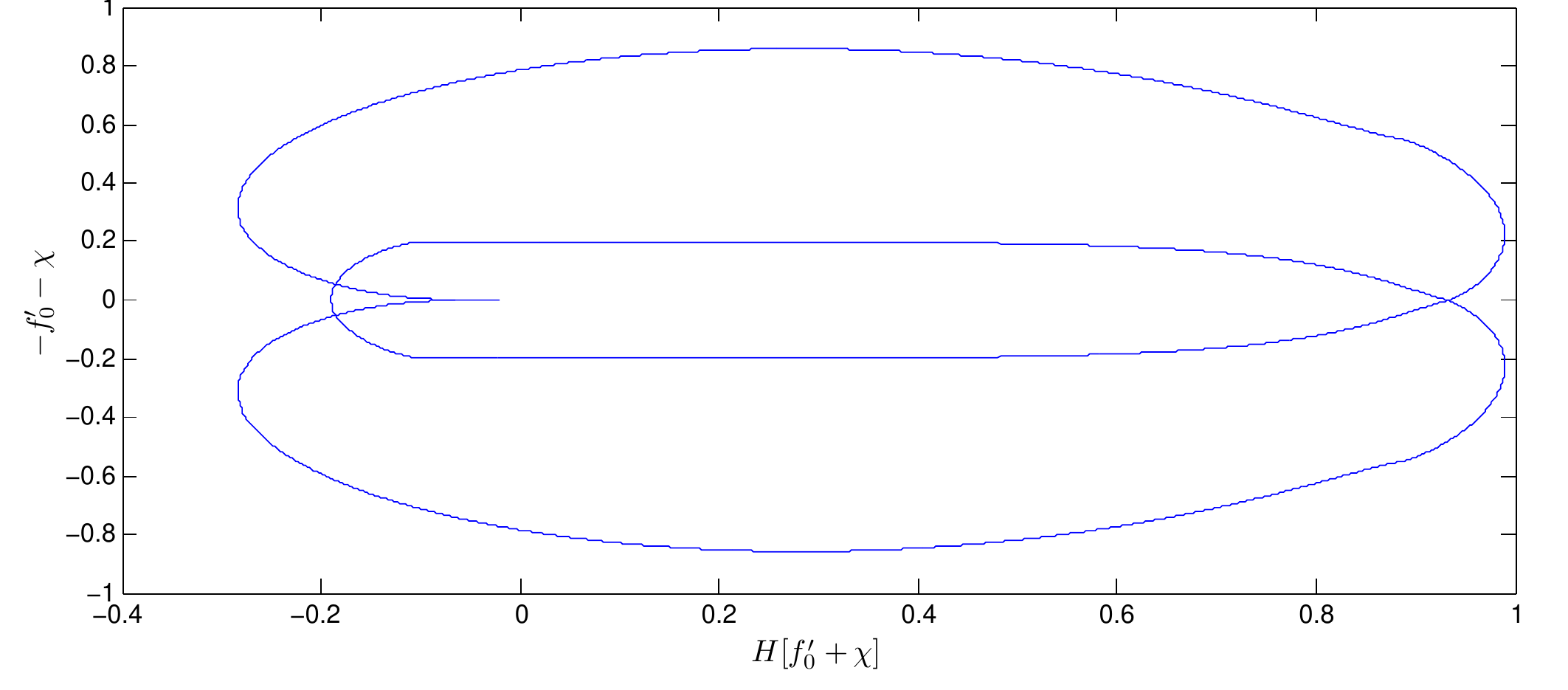}
\end{center}
\caption{Penrose plot for perturbed Maxwellian.}
\label{pertpen}
\end{figure}

Thus we emphasize that we can always construct a perturbation
that makes our linearized Vlasov-Poisson system unstable.  
For the special case of the Maxwellian distribution,  Fig.~\ref{pert} shows the perturbed derivative of the distribution function and  Fig.~\ref{pertpen} shows the Penrose plot of the unstable perturbed system. Observe the two crossings created by the perturbation  on the positive axis as well as the negative crossing arising from the unboundedness of the perturbation. 

In some sense Theorem \ref{structuralinstability}   represents a failure
of our class of perturbations to produce any interesting structure for the Vlasov
equation.  Indeed   signature appears to play no role in delineating bifurcation to instability.  In order to derive a nontrivial result we   develop a new theory analogous to the finite-dimensional Hamiltonian perturbation theory developed by Krein
and Moser.  This new theory involves a restriction  to dynamically accessible perturbations of the equilibrium state.  This is natural since the noncanonical Hamiltonian structure can be viewed as the  union of canonical Hamiltonian motions (on symplectic leaves) labelled  by the equilibrium state -- to compare with traditional finite-dimensional theory requires restriction to the given canonical Hamiltonian motion under consideration. 

\bigskip

\section{The Krein-Moser theorem}
\label{secKM}

For linear  finite-dimensional Hamiltonian systems,  Hamilton's equations are a set
of first order linear ODEs. If the Hamiltonian is time-independent, then 
the behavior of solutions is   characterized by the eigenfrequencies.
If all of the eigenfrequencies are on the real axis and nondegenerate, then the system will be stable. 
If there are degenerate eigenvalues the system will be stable as long as the time evolution operator
does not have any nontrivial Jordan blocks, but there will be secular growth if it does. Any complex
eigenfrequencies will lead to instability. The Hamiltonian of a linear finite-dimensional Hamiltonian system
is a quadratic form in the canonical variables. If we consider perturbations of the
coefficients of the quadratic form it is trivial to define a notion of small perturbations,
as the resulting perturbation of the Hamiltonian will be a bounded operator. Krein and Moser
independently proved a theorem characterizing the structural stability of these systems in
terms of a signature, a quantity that amounts to the  sign of the energy evaluated on the eigenvector of a  mode.  
We consider a special case of their theorem. 
\begin{theorem}{\rm (Krein-Moser)}
\label{KreinMoser}
Let $H$ define a stable linear finite-dimensional Hamiltonian system. Then $H$ is structurally
stable if all the eigenfrequencies are nondegenerate. If there are any degeneracies, $H$ is
structurally stable if the assosciated eigenmodes have energy of the same sign. Otherwise $H$ is
structurally unstable.
\end{theorem}

This Krein-Moser theorem gives a clear picture of the behavior of these systems under
small perturbations. The eigenfrequencies move around, but remain confined to the real
line unless there is a collision between a positive energy and negative energy mode, in which case they may leave the axis.
This theorem was first proved by Krein in the early 1950s and later rediscovered by
Moser in the late 1950s. Our goal is  to place the perturbation theory of infinite-dimensional
Hamiltonian systems in the language of the finite-dimensional theory.

The appropriate definition of signature for the continuous spectrum of the Vlasov-Poisson
equation was introduced in \cite{MP92,morrison00} (see also \cite{MS08}), where an integral transform was also introduced for constructing  a canonical transformation to action-angle variables for the infinite-dimensional system.  The transformation is a generalization
of the Hilbert transform and it can be used to show that the linearized Vlasov-Poisson
equation is equivalent to the system with the following Hamiltonian functional:
\begin{align}
&H_{L}=\sum_{k=1}^{\infty}\int_{\mathbb{R}}\!du\, \sigma_k(u)\omega_k(u)J_k(u,t)\,,
\end{align}
where $\omega_k(u)=|ku|$ and $\sigma_k(u)=-\sgn(kuf_0'(u))$ is the analog of the Krein signature corresponding
to the mode labeled by $u\in\R$.   (Note, the transformation can always be carried out in a frame where  $f_0'(0)=0$.  Because the Hamiltonian does not transform as a scalar for frame shifts, which are time dependent transformations, signature is   frame dependent. The Hamiltonian in a shifted frame is obtained by adding a constant times the momentum $P_L$ of (\ref{pl}) to $H_L$.   Later we will see that  Hamiltonians that can be made sign definite is some frame are structurally stable in a sense to be defined.)

\newtheorem{signature}{Definition}
\begin{signature}
Suppose $f_0'(0)=0$. Then the signature of  the point $u\in\mathbb{R}$ is
$-\sgn(uf_0'(u))$.
\end{signature}

Below is a graph that illustrates the signature for a bi-Maxwellian distribution
function.

\bigskip 

\begin{figure}[htbp]
%\begin{center}
\hspace{ .2 in}
\includegraphics[scale=.9]{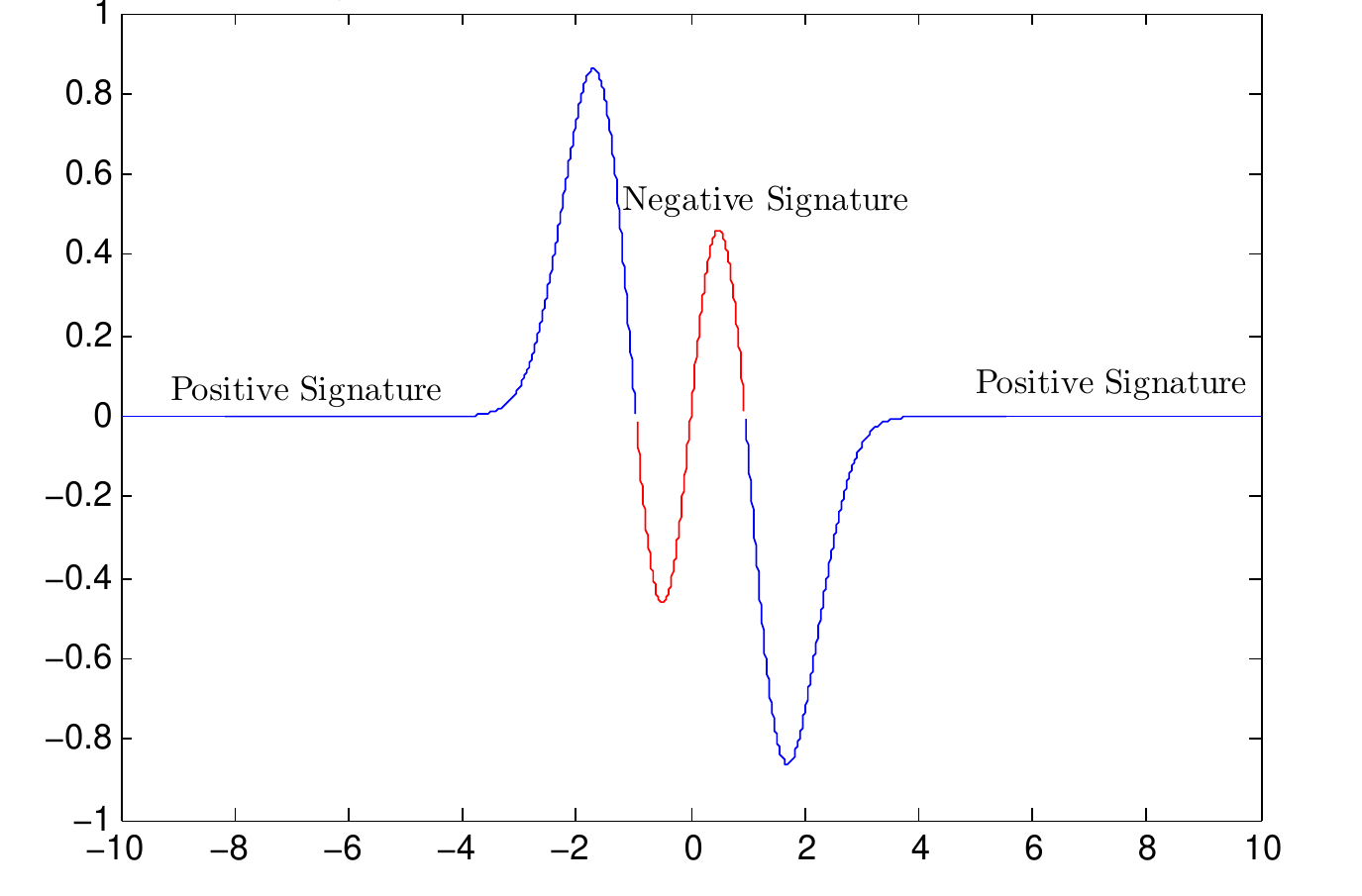}
%\end{center}
\caption{Signature for a bi-Maxwellian distribution function.}
\label{bigaus}
\end{figure}

\subsection{Dynamical accessibility and structural stability}

Now we   discuss the effect of restricting to dynamically accessible
perturbations on the structural stability of $f_0$. In this work we only study perturbations
of $f_0$ that preserve homogeneity. Because dynamically accessible perturbations are
area preserving rearrangements of $f_0$,  it is impossible to construct a dynamically accessible 
perturbation for the Vlasov equation in a finite spatial domain that preserves homogeneity.

To see this we write a rearrangement as $(x,v)\leftrightarrow (X,V)$,  where $V$ is a function of
$v$ alone.  Because  $[X,V]=1$ and  $V(v)$ is not a function of $x$,  we have 
$V'{\p X}/{\p x}=1$, or   $X={x}/{V'}$. If the spatial domain is finite, 
this map is not an diffeomorphism unless $V'=1$. 
In the infinite spatial domain case,  this is not  a problem and these rearrangments
exist. First we note that a rearrangement cannot change the critical points of $f_0$.

\begin{lemma}
\label{criticalda}
Let $(X,V)$ be an area preserving diffeomorphism, and let $V$ be homogeneous. Then the critical points
of $f_0(V)$ are the points $V^{-1}(v_c)$,  where $v_c$ is a critical point of $f_0(v)$.
\end{lemma}
\begin{proof}
By the chain rule ${df_0(V(v))}/{dv}=V(v)'f_0'(V(v))$. The function $V'\neq 0$ because $(X,V)$ must
be a diffeomorphism. Therefore the critical points occur when $f_0'(V)=0$ or at points $v=V^{-1}(v_c)$.
\qqed
\end{proof}

Consider the perturbation $\chi$ that was constructed earlier. If $v_c$ is a nondegenerate critical point of $f_0$ 
such that $f_0''(v_c)<0$, 
then we want to prove that there is a rearrangement $V$ such that $f_0(V)=f_0(v)+\int_{-\infty}^v\chi(v'-v_c)dv'$
or that ${df_0(V)}/{dv}=f_0'(v)+\chi(v-v_c)$. Such a rearrangement can be constructed as long as the parameters defining $\chi$,
the numbers $h,d,\epsilon$,
are chosen such that $f_0'(v)+\chi(v-v_c)$ has the same critical points as $f_0'(v)$. Using Morse theory it is possible to construct
a $V$
so that $f_0(V)=f_0(v)+\int\chi+O((v-v_c)^3)$, where $O((v-v_c)^3)$ has compact support and is smaller than 
$f_0(v)-{f_0''(v_c)}(v-v_c)^2/2$.

\begin{theorem}
\label{rearrangement}
Let $v_c$ be a nondegenerate critical point of $f_0$ with  $f_0'(v_c)<0$. Then there exists a rearrangement $V$ 
such that $f_0(V)=f_0(v)+\int\chi+O((v-v_c)^3$,  where $O$ is defined as above.
\end{theorem}
We omit the proof but it is a simple application of the Morse lemma. In order to apply the Morse lemma $f_0$ must be $C^2$. This is not restrictive
for practical applications where typically $f_0$ is smooth. The rearrangement of $f_0$ can also be made to be smooth if desired.

\bigskip

Using this result we prove a Krein-like theorem for dynamically accessible perturbations
in the $W^{1,1}$ norm.

\begin{theorem}
\label{kreinlike}
Let $f_0$ be a stable equilibrium distribution function for the Vlasov equation on an
infinite spatial domain. Then $f_0$ is structurally stable 
under dynamically accessible perturbations in $W^{1,1}$,  if there is only one solution 
of $f_0'(v)=0$. If there are multiple solutions,  $f_0$ is structurally unstable and the
unstable modes come from the zeros of $f_0'$ that satisfy $f_0''(v)<0$.
\end{theorem}
\begin{proof}
Suppose that $f_0'$ has only one zero on the real line. Because $f_0$ is an equilibrium this
zero will have $f_0''>0$. Because a dynamically accessible perturbation can never increase the number
of critical points,  it will be impossible to change the winding number of the Penrose plot to a positive
number. Therefore $f_0$ is structurally stable. 

Suppose that $f_0'=0$ has more than one solution on the real axis. Using the previous theorem
perturb $f_0'$ by $\chi(v-v_c)$ in a neighborhood of a critical point $v_c$ with $f_0''(v_c)<0$.
This will increase the winding number to 1 since it will add a positively oriented crossing on the 
negative real axis for the correct choices of $h$, $d$, and $\epsilon$ in the definition of $\chi$.
The norm of $\chi$ can be made as small as necessary and therefore $f_0$ is structurally unstable.
Since no other critical points with $f_0''<0$ can be created the only critical points that lead to instabilities
are the ones that already exist having $f_0''<0$.
\qqed
\end{proof}

The implication of this result is that in a Banach space where the Hilbert transform is
an unbounded operator the dynamical accessibility condition makes it so that a change in
the Krein signature of the continuous spectrum is a necessary and sufficient condition
for structural instability. The bifurcations do not occur at all points   where the 
signature changes, however. Only those that represent valleys of the distribution can 
give birth to unstable modes.

\section{Krein bifurcations in the Vlasov equation}
\label{kreinvp}

\begin{figure}[htbp]
\begin{center}
\includegraphics[scale=.6]{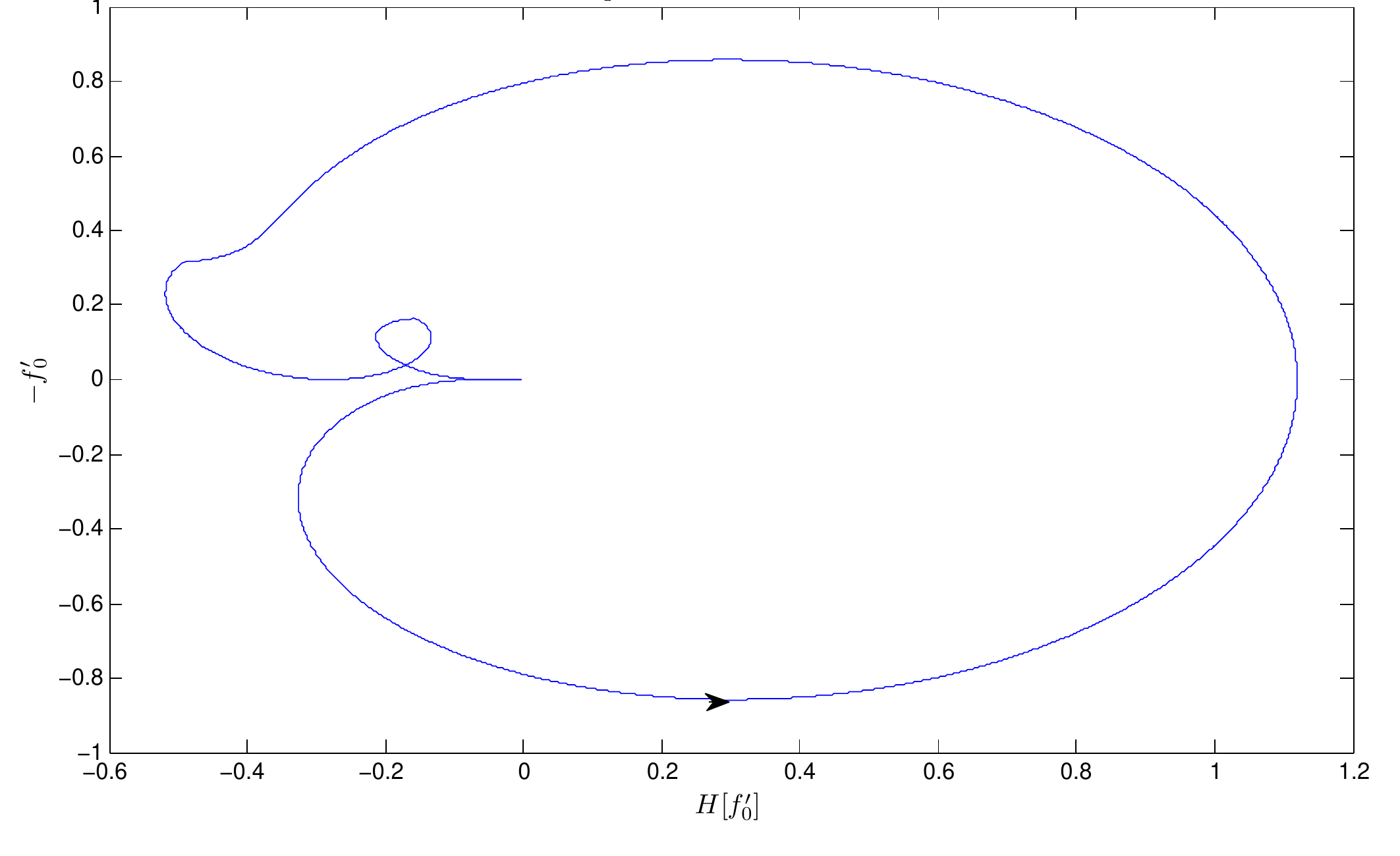}
\end{center}
\caption{Critical Penrose plot for  a $k\neq 0$ bifurcation.}
\label{gauscrit}
\end{figure}

\begin{figure}[htbp]
\begin{center}
\includegraphics[scale=.6]{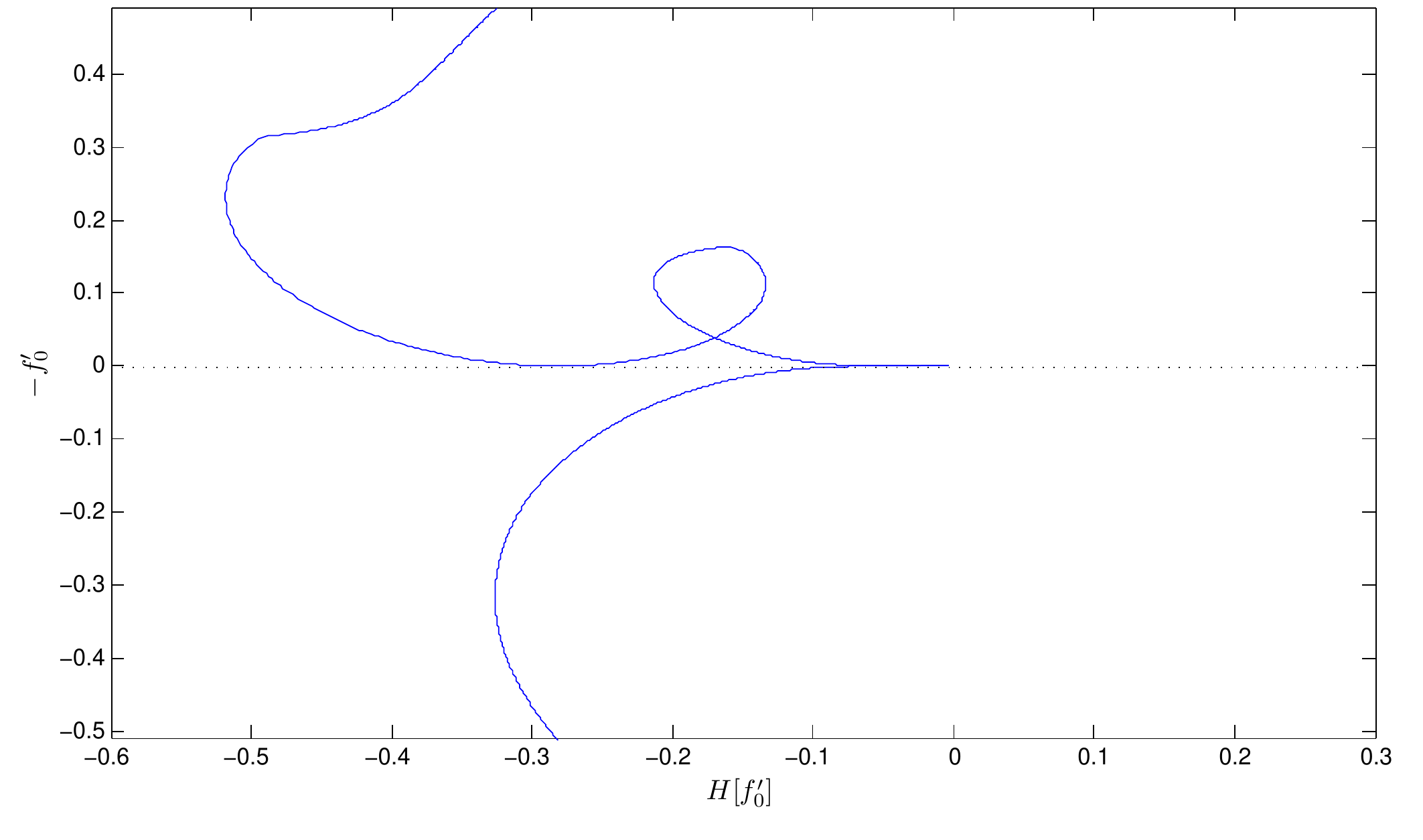}
\end{center}
\caption{Close up of a critical Penrose plot for a $k\neq 0$ bifurcation.}
\label{pengauscrit}
\end{figure}

 We identify two
critical states for the Penrose plots that correspond to the transition to instability. In these
states the system may be structurally unstable under infinitesimal perturbations of $f_0'$ in the
$C^n$ norm for all $n$.
The first critical state corresponds to the existence of an embedded mode in the continuous spectrum.
If the equilibrium is stable,  then such an embedded mode corresponds to a tangency of the Penrose plot to the real axis at
the origin.
If the system is
perturbed so that the tangency becomes a pair of transverse intersections, then  the winding number of the Penrose plot
would jump to 1 and the system would be unstable. Considering a parametrized small perturbation,  we see that the value of $k$  for the unstable mode, will correspond to some  value of $k\neq 0$ for which the embedded mode exists.  Figures \ref{gauscrit}  and \ref{pengauscrit}  illustrate a critical Penrose plot for a bifurcation at $k\neq 0$.  We explore this bifurcation in Sec.~\ref{kne0}.

\begin{figure}[htbp]
\begin{center}
\includegraphics[scale=.6]{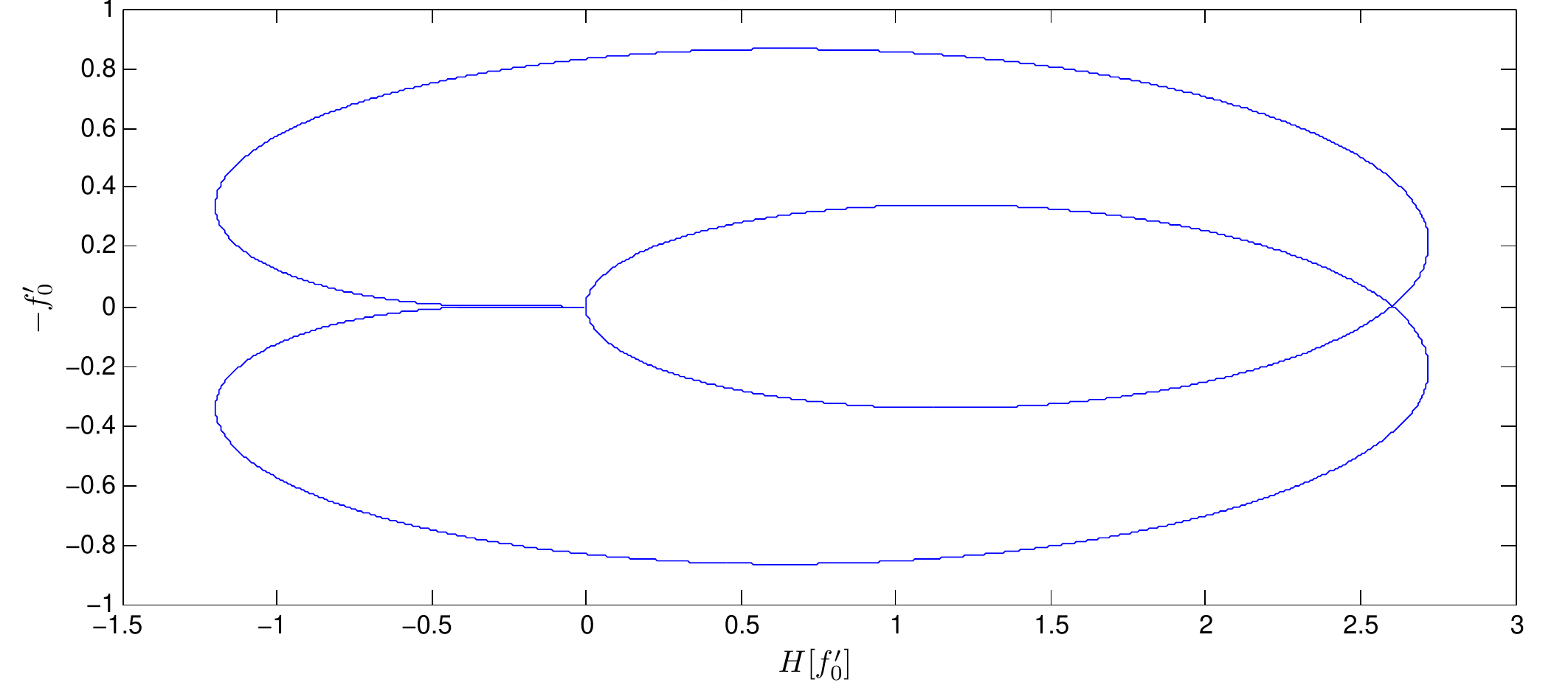}
\end{center}
\caption{Critical Penrose plot for a bi-Maxwellian distribution function.}
\label{bigaus2}
\end{figure}

Another critical state occurs when $H[f_0']=0$ at a point
where $f_0'$ transversely intersects the real axis.  
If the Hilbert transform of $f_0'$ is perturbed,  there
will be a crossing with a negative $H[f_0']$, and  the winding number will be positive for some $k$. 
This mode enters through $k=0$ because the smaller the perturbation of $H[f_0']$
the smaller $k$ must be for $T_k$ to be unstable. Figure \ref{bigaus2} is  a critical Penrose
plot corresponding to the  bi-Maxwellian distribution with the maximum stable separation.  We explore this kind of bifurcation   in Sec.~\ref{ke0}.

\bigskip

%%%%%%%%%%%%%%%%%%%%%%%%%%%%%%%%%

%%%%%%%%%%%%%%%%%%%%%%%%%%%%%%%%%

\subsection{Bifurcation at $k\neq 0$}
\label{kne0}

The linearized Vlasov equation can support neutral plasma modes embedded within the continuous spectrum.
The condition for existence of a point mode is the vanishing of the plasma dispersion
relation on the real axis,
\begin{align}
\vep(u) &= 1-H[f_0']+if_0'= 0\,.
\end{align}
If the spatial domain is unbounded the point modes will be analogues of 
the momentum eigenstate solutions of the Schrodinger equation and have infinite energy.
Any violation of the Penrose criterion will guarantee the existence
of zeros of the plasma dispersion function on the real axis because $k$ can take any 
value in this case.

If the plasma dispersion vanishes at some $u$ and $f_0''(u)=0$,  there is an embedded mode in 
the continuous spectrum. The signature of the continuous spectrum will not change signs at
the frequency of the mode and we will extend the definition of signature to the point $u$ even though
$f_0'(u)=0$. 
The signature of an embedded mode is given by $\sgn(u\,{\p\epsilon_R}/{\p u})$ (see \cite{MP92,SM94}).   
The signature of the continuous spectrum is $-\sgn(uf_0')$.
These  signatures are the same if  the value of $f_0'$ in a neighborhood of its zero is the same sign as
$H[f_0'']$.

We will prove that if $f_0$ is stable and mildly regular, it is impossible for there to be a discrete mode embedded in
the continuous spectrum with signature that is the same as the signature of the continuous
spectrum surrounding it. The proof has a simple conceptual outline. Suppose that there exists
a discrete mode with the same signature as the continuum. Then there exists some point
$u$ satisfying $f_0'(u)=0$, $-\sgn( f_0')=\sgn({\p \epsilon_R}/{\p u})$ in a 
neighborhood of $u$, and $H[f_0'](u)=0$. Perturbations of $f_0'$ centered around this
point will give the Penrose plot a negative winding number, contradicting the analyticity
of the plasma dispersion function in the upper half plane. We need $f_0'$ to be Holder continuous
so that the Penrose plot is continuous and for the plasma dispersion function to converge uniformly to
its values on the real line.

\begin{lemma}
Let $g$ be a function defined on the real line such that $g$ is Holder and
let $h=H[f]$. Then the functions $g_z$, $h_z$ that are the solutions of the Laplace equation
in the upper half plane satisfying $f_z=f$ and $g_z=g$ on the real line converge 
uniformly to $f$ and $g$.
\end{lemma}
\label{firstlemma}
\begin{proof}
Because $g$ can be defined as a bounded and continuous function on  $\mathbb{R}\cup\{\infty\}$ 
and the $g_z$ are analytic,  the $g_z$
must converge uniformly to $g$. The same properties hold for $h$  and $h_z$ must converge to
$h=H[g]$.
\qqed
\end{proof}
\begin{lemma}
Let $f_0'$ be the derivative of an equilibrium distribution function and let $f_0'$ 
be sufficiently regular such that the assumptions of the previous lemma are true.
Then the Penrose plot that is associated with $f_0'$ cannot have a negative winding number.
\end{lemma}
\label{previouslemma}
\begin{proof}
The Penrose plot associated with $f_0'$ is the image of the real line
under the map $\epsilon(u)=1-H[f_0']+if_0'$. 
This is naturally defined as an analytic function if $u$ is in the upper half plane. By the argument principle the image of 
$\mathbb{R}+it$ under
this map has a non-negative winding number. Both the
real and imaginary parts of this map converge uniformly to their values on the real line. Therefore the Penrose plot 
is a
homotopy of these contours, making it possible to parametrize the contours by some $t$ such that the
distance from the Penrose plot to the contour produced by the image of $\mathbb{R}+i\delta$ is always
less than some $\eta(t)$ that goes to $0$. If the winding number of the Penrose plot were negative,  there
would be some $t$ for which the winding number was negative because the winding 
is a stable property under homotopy, contradicting the analyticity of the map.
\qqed
\end{proof}

\begin{theorem}
Let $f_0'$ and $f_0''$ be Holder continuous. If $f_0$ is stable there are
no discrete modes with signature the same as the signature of the continuum.
\end{theorem}
\begin{proof}
Because $f_0$ is stable the winding number of the Penrose plot is equal to 0.
Assume that there is a discrete mode with the same signature as the contiuum surrounding it.
Then there exists a point $u$ with $f_0'(u)=0$, $f_0''(u)=0$,  and $\sgn(f_0'(u+\delta)dH[f_0'](u)/{du})=1$.
Then we search for a function $g$ such that the
Penrose plot of $f_0'+g$ has a negative winding number. If such a function exists it will
contradict Lemma \ref{previouslemma}. Because $f_0''$ is Holder ${\p\epsilon_R}/{\p u}$ is bounded 
away from zero in a neighborhood of the point. Suppose that in this neighborhood
there is only one zero of $f_0'$. Then define $g$ such that $g$ has one sign, is smooth and has compact
support,  and such that the $|\p {H[g]}/{\p u}|<|f_0''|$ in this 
neighborhood. Then for small enough $g$ the function $f_0'+g$ will have two zeros in a neighborhood of the point. 
Then both of the crossings
will correspond to crossings of negative orientation and the resulting winding number 
will be $-1$, a contradiction.
\qqed
\end{proof}

\begin{corollary}
\label{otherobstruction}
If $f_0$ is stable it is impossible for there to be a point where $f_0'=0$, $f_0''<0$, and $H[f_0']>0$.
\end{corollary}

If $f_0$ is unstable the winding number is positive. In this case it may be possible for modes with
the same signature as the continuum to exist. 
It is possible for a positive energy mode to be embedded in a section of negative
signature and a negative energy mode to be embedded in a section of positive signature.
This situation is structurally unstable under perturbations that are bounded by the $C^n$ norm 
and remains so even when a linear dynamical accessibility constraint is enforced.

\begin{theorem}
Let $f_0'$ be the derivative of an equilibrium distribution function with a discrete mode
embedded in the continuous spectrum. Then there exists an infinitesimal function with compact 
support in the
$C^{n}$ norm for each $n$ such that $f_0'+\delta f'$ is unstable.
\end{theorem}
\begin{proof}
Suppose that ${H[f_0'']}$ is non-zero in a neighborhood of the 
embedded mode. Define a dynamically accessible perturbation $\delta f=hf_0'$. Then assume
that $f_0'''\neq 0$ at the mode. If we define $h$ such that it does not vanish at the 
mode we find that $\delta f''=h''f_0'+h'f_0''+hf_0'''$ and therefore we can choose
$h$ such that the discrete mode becomes a crossing. This can be done with $h$ 
infinitesimal and smooth. The resulting perturbation will have an infinitesimal effect
on $f_0'$. The new crossings will cause a violation of the Penrose criterion, and
therefore the system with the embedded mode is structurally unstable.
\qqed
\end{proof}
This is an analog of Krein's theorem for the Vlasov equation for the case where there is
a discrete mode. As a result of this we see that all discrete modes are either unstable
or structurally unstable. 

\subsection{Little-big man theorem}

Consider a linearized equilibrium that supports three discrete modes. The signature
of each mode depends on the reference frame. There is a result that applies to a number
of Hamiltonian systems, the three-wave problem in particular \cite{CRS69,KM95}),  
that gives a condition on the signature of the modes and their 
frequency in some reference frame such that no frame shift can cause all the modes to
have the same signature.  In a shifted frame the Hamiltonian changes by virtue of the 
frequencies in the action-angle from being doppler shifted.  Sometimes such shifts can render the Hamiltonian sign definite.  The little-big man theorem for finite systems indicates when this cannot happen.  Such a result is also possible for the point spectrum of the Vlasov equation.

\begin{theorem}
Let $f_0'$ be the derivative of an equilibrium distribution function that has three
discrete modes (elements of the point spectrum) with real frequencies. Consider a reference frame where all of the modes
have positive frequency. Then represent the energies of the three modes as a triplet
$(\pm\pm\pm)$ where the plus and minus signs correspond to the signature of each mode,
with the first mode being the one with the lowest frequency and the last the one with
the highest frequency. Then if the triplet is of the form $(+-+)$ or $(-+-)$ there
is no reference frame in which all the modes have the same signature. If the triplet has
any other form, then  there is a reference frame in which all the modes have the same signature.
\end{theorem}

The formula for the energy of an embedded mode is $\sgn (u\,{\p \epsilon_R}/{\p u})$.
If there are three embedded modes in a frame where the frequencies are all positive the
triplet is 
\[
\left(\left.\sgn\frac{\p\epsilon_R}{\p\omega}\right|_{\omega_1},\sgn\left.\frac{\p\epsilon_R}{\p\omega}\right|_{\omega_2},
\sgn\left.\frac{\p\epsilon_R}{\p\omega}\right|_{\omega_3}\right)
\]
\begin{proof}
If this is $(+-+)$ then as we shift frames
the possible triplets are $(0-+),(--+),(-0+),(-++),(-+0),(-+-)$. All of these are
indefinite. The other possibile initially indefinite triplet is $(--+)$. However if we 
shift the two $-$ modes to negative frequency the triplet becomes $(+++)$. All other 
examples are either definite or reduce to one of these two.
\qqed
\end{proof}

\subsection{Bifurcation at $k=0$}
\label{ke0}

Assume that there are no embedded modes and that $f_0$ is stable, but that there is a point that has $f_0'=0$ and $H[f_0']=0$.
This is
the critical state for a bifurcation at $k=0$. This can be destabilized in the same way as the critical state for
 $k\neq 0$. There will be a perturbation that makes $H[f_0']<0$ without changing $f_0'$ at that point.
Therefore the Penrose plot becomes unstable and the equilibrium is structurally unstable. 

\begin{theorem}
\label{infinitek}
Suppose that $f_0'$ is a stable equilibrium distribution function that has a zero at $u$
of both $f_0'$ and $H[f_0']$. Then $f_0'$ is structurally unstable under perturbations 
bounded by the $C^n$ norm for all $n$.
\end{theorem}

\begin{proof}
Let $\delta h$ be symmetric about the point $u$, be smooth with compact support and have its first $n$ derivatives less than 
some $\epsilon$. Then let $\delta f_0'=-H[\delta h]$. The resulting
perturbation to $H[f_0']$ is $h$. If $h$ is positive at $u$, then by the symmetry of $h$
$f_0'+\delta f_0'$ has a zero at $u$ and $H[f_0']+h$ is positive there. Thus the Penrose plot has a positive
winding number and is unstable. Therefore $f_0'$ is structurally unstable.
\qqed
\end{proof}

The previous two sections demonstrated that when the Penrose plot is critical,  no amount of regularity
is sufficient to prevent $f_0$ from being structurally unstable. However,  when the Penrose plot is not 
critical all that is required is that a small perturbation only change the Penrose plot by a small amount
in addition to a condition to prevent perturbations near $v=\infty$. Suppose we arbitrarily restrict the support
of the perturbations so that $|v|<v_{max}$. Then if we increase the required regularity such that ${\rm sup} (H[\delta f_0'])$
is bounded there will be some $\delta$ such that for all $\delta f_0'$ with $\|\delta f'_0\|<\delta$ the distribution
$f_0+\delta f_0$ is structurally stable. This restriction can be motivated physically by restricting the particles in the
distribution function to be travelling slower than the speed of light.

%%%%%%%%%%%%%%%%%%%%%%%%%%%%%%%%%

%%%%%%%%%%%%%%%%%%%%%%%%%%%%%%%%%

\section{Conclusion}
\label{conclu}

We have considered perturbations of the linearized Vlasov-Poisson equation through
changes in the equilibrium function. The effect of these perturbations on the spectral 
stability of the equations is determined by the class of allowable perturbations and the signature of
the contiuous and point spectra. Every equilibrium can be made unstable by adding an arbitrarily small
function from the space $W^{1,1}$. If we rearrange $f_0$ then only when the signature of $f_0$ changes
sign can an arbitrarily small perturbation destabilize it. When $f_0$ is stable discrete modes 
always have the opposite signature of the spectrum surrounding them. The equilibria are structurally unstable
under $C^n$ small perturbations for all $n$. The signature of the spectrum and the signature of the discrete mode can
never be the same.

This generalization of Krein's theorem is  more complicated than the finite-dimensional original. However the
basic ideas of Krein's theorem are still important in the infinite-dimensional case. When the perturbations 
are more restricted than just belonging to $W^{1,1}$ the structural stability is determined by the signature of the spectrum.
Just as in Krein's theorem there must be a positive signature interacting with a negative signature to produce structural 
instability.

This paper was devoted primarily to the Vlasov equation, but  other noncanonical Hamiltonian systems admit to a similar
treatment, e.g.\  the 2D Euler equation with shear flow equilibria, and we hope  to chronicle  such cases in future publications.

\subsection*{Acknowledgments}

This work was supported   by the U.S.\  Dept.~of Energy
Contract No.~DE-FG03-96ER-54346.

%%%%%%%%%%%%%%%%%%%%%%%%%%%%%%%%%%%%%%%%%%%%%%%%%

%%%%%%%%%%%%%%%%%%%%%%%%%%%%%%%%%%%%%%%%%%%%%%%%%
\bibliographystyle{unsrt}

\bibliography{krein}

\begin{thebibliography}{10}

\bibitem{MP92}
P.~J. Morrison and D.~Pfirsch.
\newblock Dielectric energy versus plasma energy, and action-angle variables
  for the {V}lasov equation.
\newblock {\em Phys.\ Fluids}, 4B:3038--3057, 1992.

\bibitem{morrison00}
P.~J. Morrison.
\newblock Hamiltonian description of {V}lasov dynamics: Action-angle variables
  for the continuous spectrum.
\newblock {\em Trans. Theory and Stat. Phys.}, 29:397--414, 2000.

\bibitem{rayleigh}
J.~W.~S. Rayleigh.
\newblock {\em The Theory of Sound}.
\newblock Macmillan, London, 1896.

\bibitem{rellich}
F.~Rellich.
\newblock {\em Perturbation Theory of Eigenvalue Problems}.
\newblock Gordon and Breach Scientific Publishers, New York, NY, 1969.

\bibitem{friedrichs}
K.~O. Friedrichs.
\newblock {\em Perturbation of Spectra in Hilbert Space}.
\newblock American Mathematical Society, Providence, RI, 1965.

\bibitem{kato}
T.~Kato.
\newblock {\em Perturbation Theory for Linear Operators}.
\newblock Springer-Verlag, Berlin, 1966.

\bibitem{krein50}
M.~G. Kre\u{i}n.
\newblock A generalization of some investigations on linear differential
  equations with periodic coefficients.
\newblock {\em Dokl. Akad. Nauk SSSR}, 73A:445--448, 1950.

\bibitem{KJ80}
M.~G. Kre\u{i}n and V.~A. Jakubovi\v{c}.
\newblock {\em Four Papers on Ordinary Differential Equations}.
\newblock American Mathematical Society, Providence, RI, 1980.

\bibitem{moser58}
J.~Moser.
\newblock New aspects in the theory of stability of {H}amiltonian systems.
\newblock {\em Comm. Pure Appl. Math.}, 11:81--114, 1958.

\bibitem{morrison98}
P.~J. Morrison.
\newblock Hamiltonian description of the ideal fluid.
\newblock {\em Rev.\ Mod.\ Phys.}, 70(2):467--521, 1998.

\bibitem{morrison05}
P.~J. Morrison.
\newblock Hamiltonian and action principle formulations of plasma physics.
\newblock {\em Phys. Plasmas}, 12:058102--1--058102--13, 2005.

\bibitem{grillakis}
M.~Grillakis.
\newblock Analysis of the linearization around a critical point of an infinite
  dimensional {H}amiltonian system.
\newblock {\em Comm. Pure. Appl. Math.}, 43:299--333, 1990.

\bibitem{mackay}
R.~S. MacKay and P.~G. Saffman.
\newblock Stability of water waves.
\newblock {\em Proc. R. Soc. Lond. A}, 406:115--125, 1986.

\bibitem{KM95}
C.~S. Kueny and P.~J. Morrison.
\newblock Nonlinear instability and chaos in plasma wave-wave interactions.
  {I}. {I}ntroduction.
\newblock {\em Phys. Plasmas}, 2:1926--1940, 1995.

\bibitem{hirota}
M.~Hirota and Y.~Fukumoto.
\newblock Energy of hydrodynamic and magnetohydrodynamic waves with point and
  continuous spectra.
\newblock {\em J. Math. Phys}, 49:083101--1--27, 2008.

\bibitem{morrison80}
P.~J. Morrison.
\newblock The {M}axwell-{V}lasov equations as a continuous {H}amiltonian
  system,.
\newblock {\em Phys.\ Lett.}, 80:383--386, 1980.

\bibitem{morrison94}
P.~J. Morrison.
\newblock The energy of perturbations of {V}lasov plasmas.
\newblock {\em Phys. Plasmas}, 1:1447--1451, 1994.

\bibitem{MS94}
P.~J. Morrison and B.~Shadwick.
\newblock Canonization and diagonalization of an infinite dimensional
  noncanonical {H}amiltonian system: Linear {V}lasov theory.
\newblock {\em Acta Phys. Pol.}, 85:759--769, 1994.

\bibitem{BM98}
N.~J. Balmforth and P.~J. Morrison.
\newblock A necessary and sufficient instability condition for inviscid shear
  flow.
\newblock {\em Studies in Appl. Math.}, 102:309--344, 1998.

\bibitem{BM02}
N.~J. Balmforth and P.~J. Morrison.
\newblock {H}amiltonian description of shear flow.
\newblock In J.~Norbury and I.~Roulstone, editors, {\em Large-Scale
  Atmosphere-Ocean Dynamics II}, pages 117--142. Cambridge, Cambridge, 2002.

\bibitem{morrison03}
P.~J. Morrison.
\newblock Hamiltonian description of fluid and plasma systems with continuous
  spectra.
\newblock In O.~U. Velasco~Fuentes, J.~Sheinbaum, and J.~Ochoa, editors, {\em
  Nonlinear Processes in Geophysical Fluid Dynamics}, pages 53--69. Kluwer,
  Dordrecht, 2003.

\bibitem{morrison82}
P.~J. Morrison.
\newblock {P}oisson {B}rackets for {F}luids and {P}lasmas.
\newblock In M.~Tabor and Y.~Treve, editors, {\em {M}athematical {M}ethods in
  {H}ydrodynamics and {I}ntegrability in {D}ynamical {S}ystems}, volume~88,
  pages 13--46. Am.\ Inst.\ Phys., New York, 1982.

\bibitem{holm85}
D.~D. Holm, J.~E. Marsden, R.~Ratiu, and A.~Weinstein.
\newblock Nonlinear stability of fluid and plasma equilibria.
\newblock {\em Phys. Rep.}, 123:1--116, 1985.

\bibitem{morrison87}
P.~J. Morrison.
\newblock Variational principle and stability of nonmonotonic
  {V}lasov-{P}oisson equilibria.
\newblock {\em Zeitschrift f. Naturforschung}, 42a:115--123, 1987.

\bibitem{SM94}
B.~Shadwick and P.~J. Morrison.
\newblock On neutral plasma oscillations.
\newblock {\em Phys. Lett.}, 184A:277--282, 1994.

\bibitem{MS08}
P.~J. Morrison and B.~Shadwick.
\newblock On the fluctuation spectrum of plasma.
\newblock {\em Comm. Nonlinear Sci. and Num. Simulations}, 13:130--140, 2008.

\bibitem{KO58}
M.~D. Kruskal and C.~Oberman.
\newblock On the stability of plasma in static equilibrium.
\newblock {\em Phys.\ Fluids}, 1:275--280, 1958.

\bibitem{MP89}
P.~J. Morrison and D.~Pfirsch.
\newblock Free energy expressions for {V}lasov-{M}axwell equilibria.
\newblock {\em Phys. Rev.}, 40A:3898--3910, 1989.

\bibitem{MP90}
P.~J. Morrison and D.~Pfirsch.
\newblock The free energy of {M}axwell-{V}lasov equilibria.
\newblock {\em Phys.\ Fluids}, 2B:1105--1113, 1990.

\bibitem{degond86}
P.~Degond.
\newblock Spectral theory of the linearized {V}lasov-{P}oisson equation.
\newblock {\em Trans. Am. Math. Soc.}, 294:435--453, 1986.

\bibitem{penrose}
O.~Penrose.
\newblock Electrostatic instabilities of a uniform non-maxwellian plasma.
\newblock {\em Phys. Fluids}, 3:258--265, 1960.

\bibitem{king}
F.~W. King.
\newblock {\em Hilbert Transforms}.
\newblock Cambridge University Press, Cambridge, 2009.

\bibitem{adams}
R.~A. Adams.
\newblock {\em Sobolev Spaces}.
\newblock Elsevier Science Ltd., Kidlington, Oxford, UK, 2003.

\bibitem{guillemin}
V.~Guillemin and A.~Pollack.
\newblock {\em Differential Topology}.
\newblock Prentice Hall, 1974.

\bibitem{CRS69}
B.~Coppi, M.~N. Rosenbluth, and R.~N. Sudan.
\newblock Nonlinear interactions of positive and negative energy modes in
  rarefied plasmas ({I}).
\newblock {\em Ann. Phys.}, 55:207--247, 1969.

\end{thebibliography}

\end{document}